\documentclass[usenatbib,useAMS]{mn2e}
\usepackage{graphics}
\usepackage{psfrag}
\usepackage{xspace}
\usepackage{amstext}
\usepackage{amsmath}
\usepackage{amssymb}
\usepackage{natbib}
\usepackage{aas_macros}
\usepackage{ulem}
\usepackage{graphicx}
\usepackage{longtable}
\usepackage{rotating}

\newcommand{\FigDir}[1]{./#1}

\def \kms {\ensuremath{\rm \,km\,s^{-1}}\xspace}

\def \masyr {\ensuremath{\rm{mas}\,\rm{yr}^{-1}}\xspace}
\def \ayr {\ensuremath{\rm{arcsec}\,\rm{yr}^{-1}}\xspace}

\def \deg {\ensuremath{^{\circ}}\xspace}

\def \um {\ensuremath{\mu {\rm m}}\xspace}

\newcommand{\ImJ}{\ensuremath {I - J}\xspace}

\newcommand{\VmI}{\ensuremath {V-I}\xspace}

\newcommand{\hstrut}[1]{\rule[5pt]{#1}{0pt}}

\newlength{\voff}
\newcommand{\ppm}{\ensuremath{\pm}}

\newcommand{\HS}[1][1.]{\hspace{\stretch{#1}}}

\begin{document}

\title[OGLE-II High Proper Motion Stars]{OGLE-II High Proper Motion Stars towards the Galactic centre}

\author[Rattenbury, N.~J. \& Mao, S.]
{Nicholas J. Rattenbury$^{1}$ and Shude Mao$^{1}$\thanks{e-mail: (nicholas.rattenbury,shude.mao)@manchester.ac.uk}\\
$^1$ Jodrell Bank Centre for Astrophysics, Alan Turing Building, The University of Manchester, Manchester, M13 9PL, UK \\
}
\date{Accepted ........
      Received .......;
      in original form ......}

\pubyear{2007}

\maketitle
\begin{abstract}
The photometry data base of the second phase of the OGLE microlensing experiment, OGLE-II, is a rich source of information about the kinematics and structure of the Galaxy. In this work we use the OGLE-II proper motion catalogue to identify candidate stars which have high proper motions.  521 stars with proper motion $\mu>50$ \masyr in the OGLE-II proper motion catalogue \citep{2004MNRAS.348.1439S} were cross-identified with  stars in the MACHO high proper motion catalogue, and the DENIS and 2MASS infra-red photometry catalogues. Photometric distances were computed for stars with colours consistent with G/K and M type stars. 6 stars were newly identified as possible nearby ($< 50$ pc) M dwarfs. 
\end{abstract}


\section{Introduction}
\label{sec:intro}

Several microlensing collaborations have carried out survey observations of many square degrees towards the Galactic centre, including the EROS \citep{1993Natur.365..623A}, MACHO \citep{2000ApJ...541..734A}, MOA \citep{2001MNRAS.327..868B, 2003ApJ...591..204S} and OGLE \citep{2000AcA....50....1U} groups. As a result of these extended survey campaigns, large data bases were generated, containing the photometry of millions of stars in and towards the Galactic centre. These data bases can be used to investigate the structure, kinematics and stellar populations of the central regions of the Galaxy. For instance, red clump giants in the OGLE-II photometry database  were used to constrain the shape and orientation of the Galactic bar \citep{2007MNRAS.378.1064R}. 

Many fundamental parameters of stellar astrophysics such as mass, temperature and luminosity,  are derived from the observation of stars in the solar neighbourhood. Our understanding of the nature of the Galaxy is also advanced through studies of the kinematics, chemical abundance and mass function of these nearby stars. A census of nearby stars has not been achieved however, with estimates of the sample deficit ranging from $\sim25$\% of stars within 10 pc \citep{2003AJ....125..354R} to $\gtrsim 30$\% \citep{1997AJ....114..388H}.

We select high proper motion stars from the OGLE-II proper motion data base  \citep{2004MNRAS.348.1439S} and cross-identify them with known high proper motion stars from the MACHO high proper motion catalogue \citep{2001ApJ...562..337A}. We obtain photometric distance estimates for OGLE-II high proper motion stars which have corresponding entries in the {\it DEep Near-Infrared Survey} of the southern sky (DENIS)  \citep{1997Msngr..87...27E,2005yCat....102002T}. In Section 2 we describe the OGLE-II proper motion data base and in Section 3 we describe how the high proper motion candidate stars were selected. Section 4 describes how the cross-identification was made between the OGLE-II high proper motion star candidates and entries in the DENIS data base. In Section 5 we produce photometric distance estimates for OGLE-II stars cross-identified with DENIS stars. Stars with photometric distances $<50$ pc are cross-identified with entries in the {\it Two-Micron All Sky Survey} (2MASS) catalogue and compared to a set of known cool dwarfs in Section 6. Section 7 contains a discussion on our results.

\section{OGLE-II Proper Motion Catalogue}
\label{sec:data}

During the second phase of the OGLE experiment {(OGLE~-~II)}, between 1997 and 2000, the Galactic centre was observed in 49 fields using the 1.3m Warsaw telescope at the Las Campanas Observatory, Chile. Each field was $0.24\deg \times 0.95\deg$ in size, the median seeing over the almost 4 year baseline was $\sim 1.3$ arcsec. The OGLE-II fields extend across the central regions of the Galactic bulge with $-11\deg < l < +11\deg$ and $-6\deg< b < +3\deg$, see Figure~\ref{fig:oglemacho}. Further details of the telescope, camera and operations can be found in \citet{1997AcA....47..319U}.  

\citet{2004MNRAS.348.1439S} determined the proper motions for millions of stars in the OGLE-II fields. The catalogue contains proper motion data for all stars with magnitudes in the range $11 \leq I \leq 18$, and  includes proper motions up to $\mu = 500$ \masyr, with an accuracy of better than 1 \masyr for stars with $12<I<14$. 

The catalogue of proper motions of stars in the OGLE-II fields \citep{2004MNRAS.348.1439S} extended the work of \citet{2003MNRAS.340.1346S} where the streaming motion of stars around the Galactic bar  was observed using OGLE-II proper motions in one field, as predicted by \citet{2002MNRAS.337..895M}. Other work which uses the full proper motion catalogue includes the comparison of proper motion dispersions to  model predictions \citep{2007MNRAS.378.1165R}.

\section{High proper motion (HPM) sample}
\label{sec:sample}

Candidate high proper motion stars are selected using the following criteria:

\begin{equation*}
\mu > 50 \,\, \masyr \hspace{1cm} \sigma_{\mu} / \mu < 0.1
\end{equation*}
where $\mu$ and $\sigma_{\mu}$ are the star proper motion and error, respectively. Out of a total of 5080236 stars in the OGLE-II proper motion catalogue, 18907 have proper motion $\mu > 50$ \masyr. Approximately $5\%$ of stars have $\sigma_{\mu} / \mu < 0.1$. 521 stars satisfy both the above criteria and form the set of high proper motion candidates for this work. A proper motion of $50$ \masyr at 100 pc corresponds to $\simeq 24$ \kms.

\subsection{Cross-identification with MACHO HPM stars}
\begin{figure}

\begin{center}
\hspace{-1cm}
\centering\includegraphics[width=1.0\hsize]{\FigDir{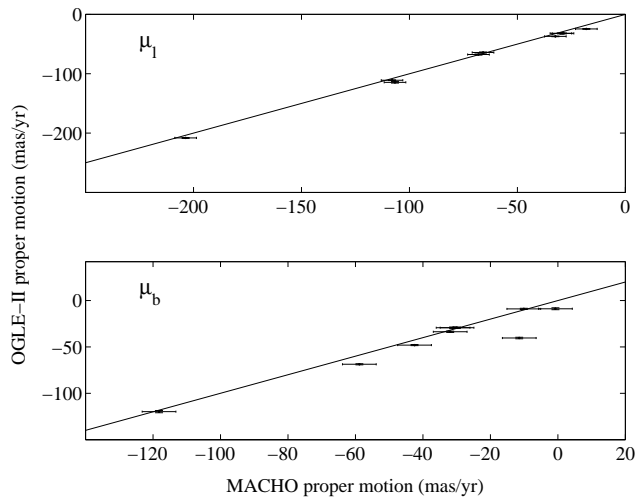}}
\end{center}
\caption{Comparison between cross-identified HPM stars in the MACHO and OGLE-II proper motion catalogues. The proper motion in the Galactic longitude and latitude directions, $\mu_{\rm l}$ (top) and $\mu_{\rm b}$ (bottom) respectively, are shown along with the line indicating $\mu_{\rm MACHO} = \mu_{\rm OGLE-II}$, see also Table~\ref{tab:crossID_machoogle}. The MACHO proper motion accuracy is $\sim 5$ \masyr \citep{2001ApJ...562..337A}; the errorbars on the OGLE-II proper motions are from the OGLE-II proper motion catalogue of \citet{2004MNRAS.348.1439S}.}
\label{fig:compoglemacho}

\end{figure}

\begin{figure*}

\begin{center}
\hspace{-1cm}
\centering\includegraphics[width=1.0\hsize]{\FigDir{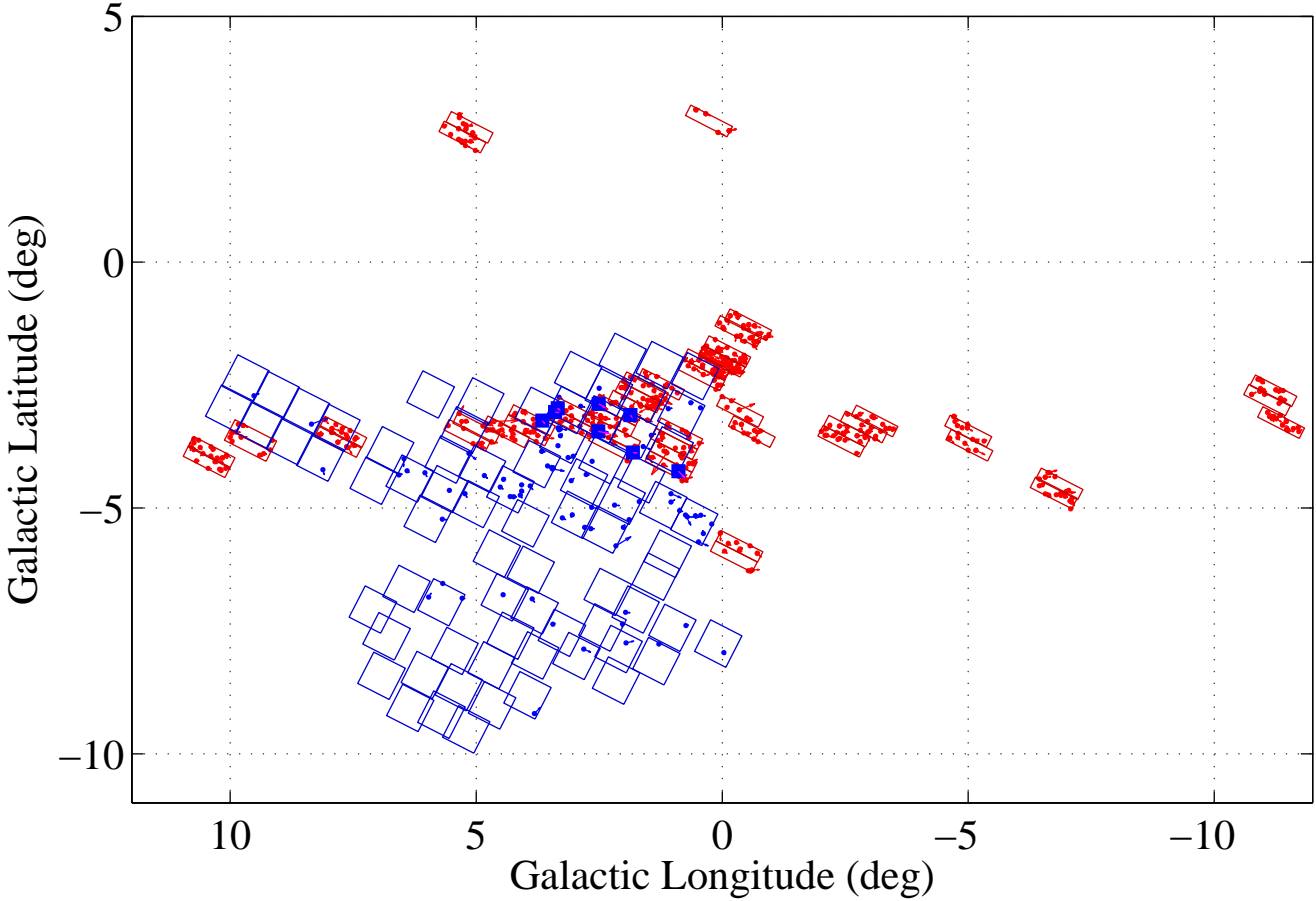}}
\end{center}
\caption{The MACHO (blue) and OGLE-II (red) Galactic bulge fields. The position and proper motion of known MACHO and candidate OGLE-II high proper motion stars are shown as blue and red points and arrows respectively. The 9 MACHO high proper motion stars which fall within the OGLE-II fields are shown as solid blue squares.}
\label{fig:oglemacho}

\end{figure*}

\citet{2001ApJ...562..337A} report 88 high proper motion stars in the MACHO Galactic bulge fields. 9 MACHO high proper motion stars are located within the OGLE-II fields. For each of these 9 stars, the nearest OGLE-II star is identified, see Table~\ref{tab:crossID_machoogle}. There is reasonable agreement between the MACHO and OGLE-II proper motion data, see Figure~\ref{fig:compoglemacho}: all are within $2\sigma$ of each other, except for one star. Figure~\ref{fig:oglemacho} shows the MACHO and OGLE-II fields, along with the 88 MACHO high proper motion stars and the sample of OGLE-II high proper motion star candidates. 

\begin{table*}
\caption{\label{tab:crossID_machoogle} Cross-identified MACHO high proper motion stars from \citet{2001ApJ...562..337A} with candidate stars from the OGLE-II proper motion catalogue \citep{2004MNRAS.348.1439S}. Data from the MACHO and OGLE-II data sets are given in columns 1--5 and 6--12 respectively.  $\mu_{\rm l}$ and $\mu_{\rm b}$ are the proper motions in the Galactic longitude and latitude directions respectively, measured in \masyr. F is the OGLE-II field number. $l$ and $b$ are the Galactic co-ordinates of the star and $\Delta \phi$ is the vector difference in star position between the OGLE-II and the MACHO catalogues, measured in arcsec. The accuracy of MACHO proper motions is $\sim 5$ \masyr.  \VmI and $I$ denote the OGLE-II colour and magnitude of each star. The MACHO stars 101.20908.263 and 104.20908.6076 have very similar positions and proper motions and are cross-identified with only one OGLE-II star.}

 \begin{tabular}{lc@{\hspace{5pt}}c@{\hspace{10pt}}r@{\hspace{5pt}}r@{\hspace{5pt}}c@{\hspace{5pt}}rr@{\hspace{5pt}}rc@{\hspace{5pt}}c@{\hspace{10pt}}c}
\hline
\multicolumn{5}{c}{MACHO} & \multicolumn{7}{c}{OGLE-II} \\
\HS ID \HS[1] &$l$ $(^{\circ})$ & \HS $b$ $(^{\circ})$ & \HS $\mu_{\rm l}$ \HS[0.5] &  \HS $\mu_{\rm b}$  \HS[0.3] &  \HS[0.5] F \HS[1]   &\HS[0.2] ID \HS[0.2] & \HS[0.7] $\mu_{\rm l}\pm \sigma_{\mu_{\rm l}}$ \HS[0.2] & $\mu_{\rm b}\pm \sigma_{\mu_{\rm b}}$ & $\Delta \phi$ &\HS[0.1] $V\!-\!I$\HS[1]& $I$ \\

\hline


  109.20894.99 & $2.52$ & $-3.44$ & $-203.64$ & $ -11.45$ & $ 2$ & $  783242$ &  $ -208.2\ppm    0.8$ & $  -40.4\ppm    0.8$ & $1.00$ & $6.19$ & $14.49$ \\
  101.21560.49 & $3.65$ & $-3.22$ & $ -32.40$ & $ -42.47$ & $18$ & $   51131$ &  $  -37.1\ppm    0.5$ & $  -48.0\ppm    0.6$ & $0.24$ & $1.74$ & $13.04$ \\
  109.19858.98 & $2.51$ & $-2.88$ & $-108.03$ & $ -58.82$ & $31$ & $  765538$ &  $ -111.0\ppm    0.6$ & $  -68.7\ppm    0.6$ & $0.71$ & $2.11$ & $13.98$ \\
  114.19718.31 & $1.87$ & $-3.11$ & $ -18.06$ & $ -31.92$ & $31$ & $  240515$ &  $  -24.5\ppm    0.6$ & $  -33.6\ppm    0.7$ & $0.42$ & $1.04$ & $13.19$ \\
 114.20882.707 & $1.82$ & $-3.87$ & $ -68.01$ & $  -0.72$ & $33$ & $     771$ &  $  -67.6\ppm    0.9$ & $   -8.8\ppm    1.0$ & $0.19$ & $1.19$ & $16.16$ \\
 101.20908.263 & $3.34$ & $-2.97$ & $ -28.88$ & $ -31.05$ & $35$ & $  750288$ &  $  -32.0\ppm    0.8$ & $  -29.3\ppm    0.7$ & $0.41$ & $0.86$ & $15.14$ \\
104.20908.6076 & $3.34$ & $-2.97$ & $ -29.82$ & $ -29.89$ & $35$ & $  750288$ &  $  -32.0\ppm    0.8$ & $  -29.3\ppm    0.7$ & $0.16$ & $0.86$ & $15.14$ \\
 101.21038.235 & $3.40$ & $-3.05$ & $ -65.90$ & $ -10.09$ & $36$ & $  426923$ &  $  -64.2\ppm    0.8$ & $   -8.9\ppm    0.7$ & $0.40$ & $2.37$ & $14.60$ \\
  119.20607.34 & $0.89$ & $-4.25$ & $-106.67$ & $-118.20$ & $46$ & $  188284$ &  $ -114.5\ppm    1.2$ & $ -119.8\ppm    1.2$ & $0.65$ & $2.15$ & $13.58$ \\

\hline

\end{tabular}
\end{table*}

\subsection{Reduced proper motion}

\begin{figure}
\begin{center}
\centering\includegraphics[width=1.0\hsize]{\FigDir{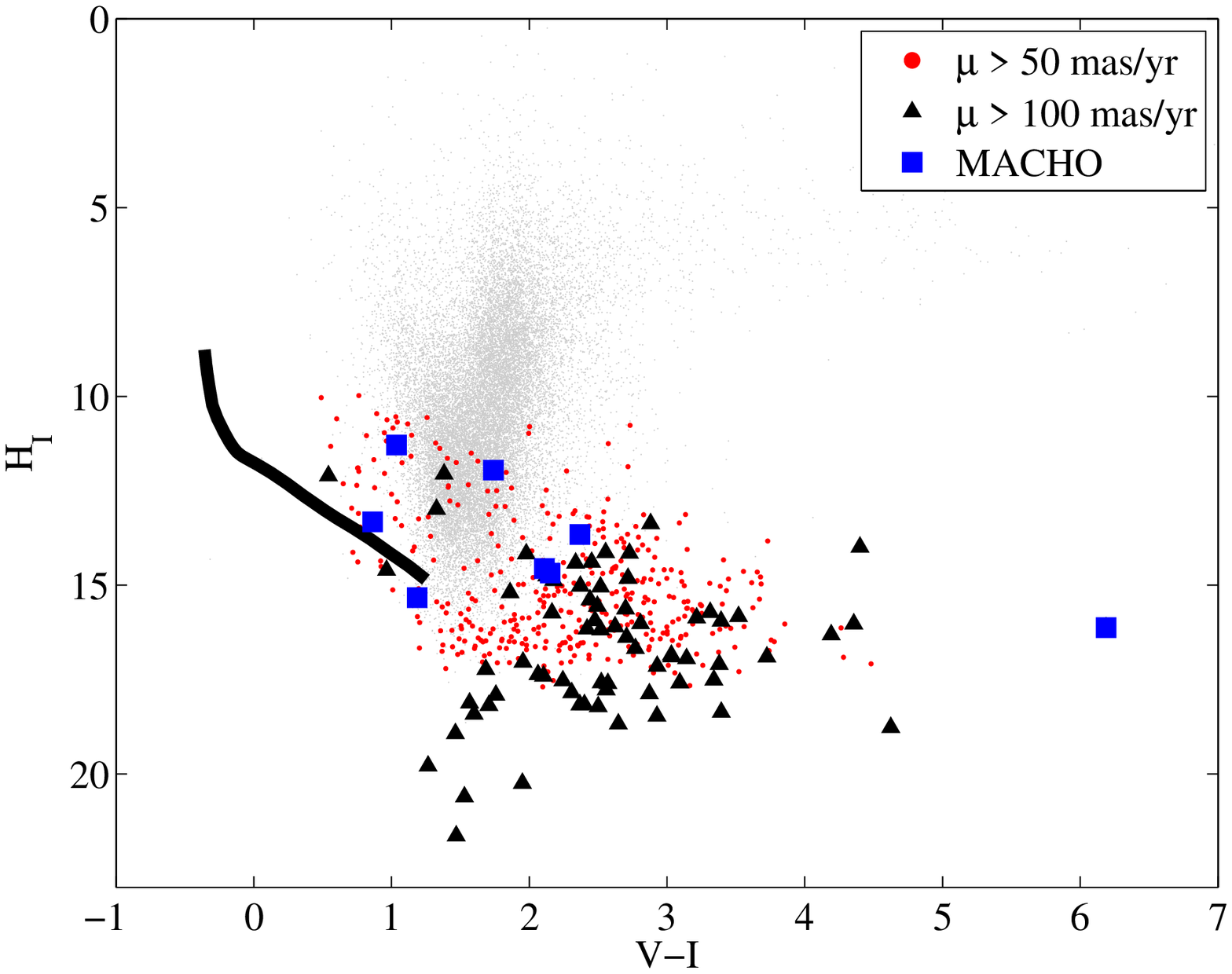}}
\end{center}
\caption{OGLE-II reduced proper motion as function of colour. The reduced proper motion for a subset of stars in OGLE bulge field SC 1 are shown in grey. Stars with proper motion $\mu>50$ \masyr or $\mu>100$ \masyr are shown as red dots and black triangles respectively. The 9 MACHO high proper motion stars which appear in the OGLE-II fields (and which are cross-identified with OGLE-II stars in Table~\ref{tab:crossID_machoogle}) are shown as solid blue squares. The solid black line is derived from the theoretical luminosity model of a pure hydrogen white dwarf from \citet{1995PASP..107.1047B}, located at a distance of 10 pc with a proper motion of 100 \masyr.}
\label{fig:reduced_ogle}
\end{figure}

We compute the reduced proper motion
\begin{equation*}
H_{\rm I} = I + 5\log\mu + 5
\end{equation*}
where $\mu$ is the proper motion in \ayr and $I$ is the OGLE-II $I$-band magnitude. Extinction is neglected as it is not expected to be a significant effect for the nearby stars of interest. Figure~\ref{fig:reduced_ogle} shows the reduced proper motion of the high proper motion candidate stars as a function of $(V\!\!-\!\!I)$. For illustrative purposes, the reduced proper motion expected from a hypothetical pure hydrogen white dwarf \citep{1995PASP..107.1047B} located at a distance of 10 pc with a proper motion of 100 \masyr is included in  Figure~\ref{fig:reduced_ogle}. A fainter or faster white dwarf would move the line relatively downwards. Similarly a brighter or slower white dwarf would move the line upward. Figure~\ref{fig:reduced_ogle} shows that most of the stars in the OGLE-II high proper motion star sample are too red to be classified as white dwarf candidates. The bluest high proper motion stars which have colours consistent with white dwarfs are more likely to be late G-type stars, rather than nearby ($< 10 $ pc) white dwarfs \citep{2002A&A...390..491R}. Infrared photometry is necessary in order to more accurately determine the stellar type of these high proper motion candidate stars.
 
\begin{figure}
\begin{center}
\centering\includegraphics[width=1.0\hsize]{\FigDir{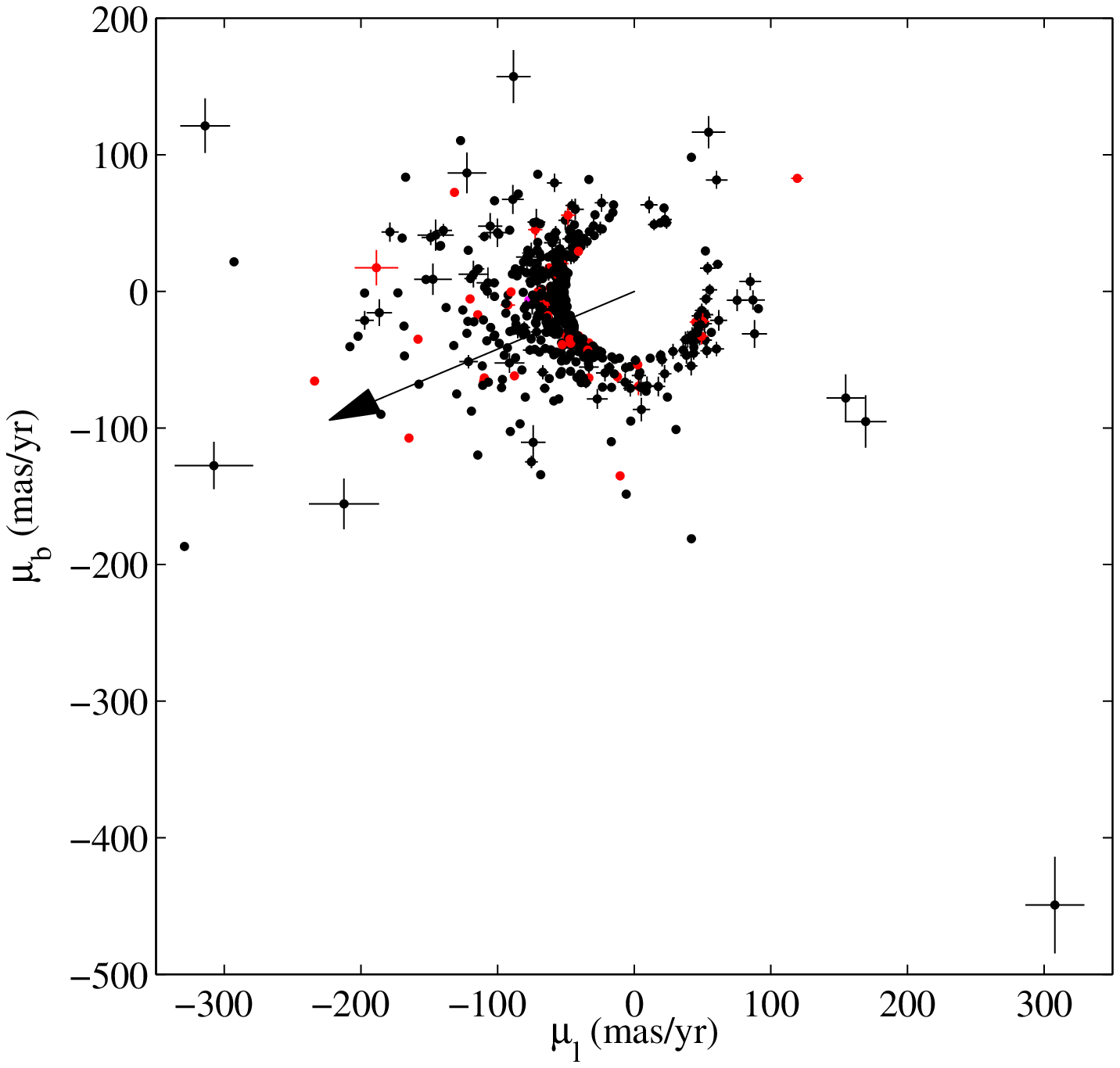}}
\end{center}
\caption{Proper motions in the Galactic co-ordinate longitude and latitude directions, $\mu_{\rm l}$, $\mu_{\rm b}$, respectively for the 521 stars in the OGLE-II proper motion catalogue which have total proper motion $\mu > 50$ \masyr and  $\sigma_{\mu} / \mu < 0.1$. The direction to the solar antapex is indicated with an arrow. Points in red correspond to stars which have photometrically determined distances of $d<50$ pc, see Section~\ref{sec:photdist}.}
\label{fig:propermotionsHPM}
\end{figure}

The proper motion in the Galactic co-ordinate longitude and latitude directions, $\mu_{\rm l}$, $\mu_{\rm b}$, respectively is shown in Figure~\ref{fig:propermotionsHPM} for the sample of 521 high proper motion stars. There is a clear preferred direction in the average motion of these stars. This is most likely to be due to the motion of the Sun with respect to the local standard of rest. The direction of the antapex is indicated in Figure~\ref{fig:propermotionsHPM}, corresponding approximately with the average proper motion. This supports the theory that many of these stars are local disk objects.

\subsection{Parallax distance determination}
The high proper motion stars are expected to display pronounced astrometric parallax signatures in their recorded position. The position of the high proper motion stars was fitted with a model similar to that of \cite{2002AcA....52..143S} which includes parallax terms. However, as noted by \citet{2001MNRAS.327..601E} and \citet{2004MNRAS.348.1439S} the influence of differential refraction is correlated with the effect of parallactic motion for stars in the Galactic bulge. A reliable parallax distance estimate for these stars was therefore not possible.

\section{Cross-identification with DENIS}
\label{sec:crossidDENIS}

The OGLE-II $I$ band filter response closely approximates the standard $BVI$ system but diverges from standard \citep{1992AJ....104..372L} magnitudes for increasingly red objects \citep{2002AcA....52..217U}. For this reason we cross-correlated OGLE-II high proper motion candidate stars with the DENIS near infra-red survey.

The DENIS survey data base contains photometry in the optical band $I$ (0.8 \um), and the near-infrared bands $J$ (1.25 \um) and $K_{\rm s}$ (2.17 \um) for the entire southern sky \citep{1997Msngr..87...27E}. The DENIS data base was searched for counterparts to stars in the sample of OGLE-II high proper motion stars. The DENIS data available online\footnote{VizieR catalogue access tool, Centre de Donn\'{e}es astronomiques de Strasbourg} use epoch J2000.0, corresponding to that used by the OGLE-II and MACHO catalogues. The OGLE-II data base has an internal positional accuracy of 0.15 -- 0.20 arcsec and is cross-identified with the {\it Digital Sky Survey} (DSS) which has a maximum systematic error of 0.7 arcsec \citep{2002AcA....52..217U}. The most pessimistic $3\sigma$ positional error for an OGLE-II star is therefore $\simeq 3$ arcsec. The DENIS frames are referenced to the USNO-A2.0 catalogue which has a $\sim 1$ arcsec accuracy \citep{2001A&A...380..590P}. The DENIS $3\sigma$ astrometric precision is also therefore $\simeq 3$ arcsec (see also \citealt{2004A&A...421..643R}). The closest DENIS star within a search radius of $(3^2 + 3^2)^{1/2} \simeq 4$ arcsec around each OGLE-II high proper motion star was determined. 241 OGLE-II high proper motion star candidates had counterparts in the DENIS database within an error circle of 4 arcsec.

\section{Photometric distances}
\label{sec:photdist}

We plot the $(I,\ImJ)$ colour-magnitude diagram for OGLE-II high proper motion candidate stars with likely DENIS counterparts in Figure~\ref{fig:denis_CMD}. Based on the range of theoretical luminosity models (see below) we identify stars with colour $\ImJ < 0.6 $ as white dwarfs; $0.6 < \ImJ < 1.0$ as G or K-dwarfs; $1 < \ImJ < 3 $  as M-dwarfs and stars with colour $\ImJ > 3.0$ as possible brown dwarf candidates. \citet{2004A&A...421..643R} note that stars with $\ImJ \lesssim 0.7 $ can be either distant red dwarfs or close white dwarfs. The theoretical luminosity-colour relations for these different stellar types are also shown in Figure~\ref{fig:denis_CMD}: for pure hydrogen white dwarfs \citep{1995PASP..107.1047B}; G/K dwarfs with metallicities [Fe/H] = 0.0, $-1.0$, $-2.0$ and $-2.5$  \citep{1997A&AS..125..229L} and M dwarfs with metallicity [Fe/H] = 0.0 \citep{1998A&A...337..403B}. These theoretical models are used to derive distance estimates for the sample of high proper motion stars. 

Tables~\ref{tab:GKstars} and \ref{tab:Mstars} show the distance estimations for the stars which are classified as either G/K dwarfs (Table~\ref{tab:GKstars}) or M dwarfs (Table~\ref{tab:Mstars}), using the corresponding theoretical luminosity models. We note that in Table~\ref{tab:GKstars} there are up to four distance estimations for stars, depending on metallicity. \citet{2002A&A...390..491R} use the Besan\c{c}on population synthesis model \citep{2003A&A...409..523R} to simulate the stellar populations in the direction of their high proper motion sample in order to discriminate between different Galactic populations. By identifying regions in the reduced proper motion-colour diagram corresponding to spheroid, thick disc and disc populations, \citet{2002A&A...390..491R} were able to apply the most appropriate metallicity value for each star. A similar clear discrimination between Galactic stellar populations using the Besan\c{c}on model in the direction of the OGLE-II fields was not possible. Distances for G/K stars derived from all applicable theoretical models are therefore presented in Table~\ref{tab:GKstars}, however we note that the metallicity for disk stars is most likely to be $-1\lesssim {\rm[Fe/H]} \lesssim 0$ \citep{1993A&A...275..101E,2002A&A...394..927I}. For this reason, more weight should be given to the first two distance estimates in column 7 of Table~\ref{tab:GKstars}. None of the stars in Table~\ref{tab:GKstars} is likely to have $d<50$ pc and are therefore not dealt with further in this study. 

One star in the OGLE-II high proper motion sample has a DENIS \ImJ colour (0.3) and I band magnitude (14.42) consistent with a close white dwarf. The photometric distance determined using the bright main branch ($I<14$ for a white dwarf at 10 pc in Figure~\ref{fig:denis_CMD}) of the theoretical luminosity model for a pure hydrogen white dwarf of \citet{1995PASP..107.1047B} is $ 18^{+   1}_{-   1}$ pc. Using the faint branch gives a distance of $  4^{+1}_{-1}$ pc. 

\begin{figure}
\begin{center}
\centering\includegraphics[width=1.0\hsize]{\FigDir{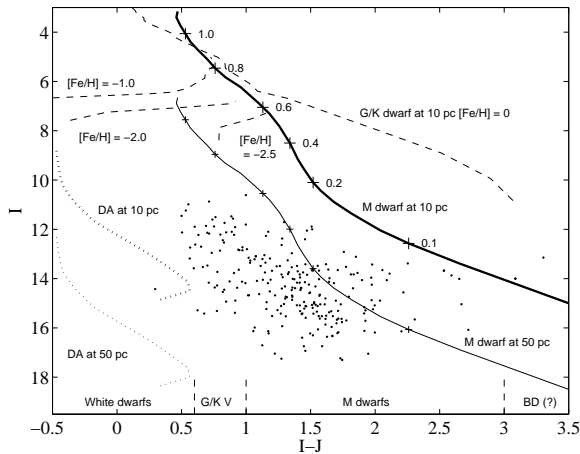}}
\end{center}
\caption{\label{fig:denis_CMD}Colour magnitude diagram of OGLE-II high proper motion stars which have DENIS counterparts. The stellar type categorization of  \citet{2002A&A...390..491R} is indicated along the lower horizontal axis, where  we identify stars with colour $\ImJ < 0.6 $ as white dwarfs;  $0.6 < \ImJ < 1.0$ as G or K-dwarfs; $1 < \ImJ < 3 $ as M-dwarfs and stars with colour $\ImJ > 3.0$ as possible brown dwarf candidates. The heavy (light) solid black line shows the luminosity-colour relation of an M-dwarf at 10 (50) pc with solar metallicity using the theoretical model of \citet{1998A&A...337..403B}. The plus signs indicate stellar mass in solar units. The theoretical luminosity-colour relations  for  a hydrogen white dwarf \citep{1995PASP..107.1047B} at 10 pc and 50 pc is shown as a dotted line. The series of dashed lines show the luminosity-colour relations for a G/K star at 10 pc with metallicities [Fe/H] = 0.0, $-1.0$, $-2.0$ and $-2.5$ \citet{1997A&AS..125..229L}.} 
\end{figure}

\section{~Search for nearby cool stars}
The 44 OGLE-II high proper motion star candidates  with distances $d<50$ pc in Table~\ref{tab:Mstars} were cross-identified with entries in the 2MASS \citep{2003tmc..book.....C} database of $JHK_{\rm s}$ photometry and are listed in Table~\ref{tab:Mstars2MASS}\footnote{Two of these stars, numbers 23 and 24 in Table~\ref{tab:Mstars2MASS} have almost identical OGLE-II positions and are treated as repeated entries.}.  37 of these have counterparts in the 2MASS database within an error circle of 1 arcsec. The photometry of bright stars in the DENIS catalogue potentially suffers effects due to saturation. We cross-identified our sample of DENIS objects with the 2MASS catalogue to obtain more accurate photometry, and to highlight instances of mis-identification. Furthermore, the 2MASS absolute astrometry accuracy is 70 - 80 mas for stars with $9< K_{\rm s} < 14$ \citep{2MASSExplSupp05}.

A total of 270 OGLE-II high proper motion star candidates had counterparts in the 2MASS database within an error circle of 3 arcsec. We note however that the DENIS $I$ magnitudes and $\ImJ$ colour were required to apply the theoretical luminosity functions above.

\section{Discussion}
\label{sec:discussion}
Table~\ref{tab:Mstars2MASS} lists 43 stars with proper motion $\mu > 50$ \masyr, distance $<50$ pc and stellar type consistent with late M dwarfs or early L type stars. The distance estimate is derived from the DENIS photometry, and the corresponding 2MASS colour and magnitude for these stars are consistent with cool dwarf stars. It is possible that a fast-moving nearby star in the OGLE-II proper motion catalogue has been falsely cross-identified with a distant red giant star in the DENIS catalogue. Categorizing this star as an M-dwarf on the basis of DENIS $\ImJ$ colour would consequently result in a small distance, and thereby $M_{J}$ values appearing to be consistent with those of late type M, or L stars. The majority of DENIS stars in Table~\ref{tab:Mstars2MASS} have a counterpart in the 2MASS catalogue within 1 arcsec. However, mis-identification is most likely to arise during the cross-identification between the OGLE-II and DENIS catalogues. We cross-identified OGLE-II stars with the DENIS catalogue entries using the most permissive ($3\sigma$) position errors in each catalogue (Section~\ref{sec:crossidDENIS}). Given the crowded nature of the fields towards the Galactic Centre, it is likely that several OGLE-II high proper motion stars have been mis-identified in the DENIS catalogue, and consequently in the 2MASS catalogue. There are 8 stars in Table~\ref{tab:Mstars2MASS} which have both the positional difference between the OGLE-II and DENIS, and between the DENIS and 2MASS positions less than 1 arcsec. Ignoring possible co-ordinate reference frame offsets between the catalogues, these 8 high proper motion stars are least likely to have been misidentified in the IR catalogues based on position alone. These stars, with Unique ID (UID) numbers 7, 9, 16, 19, 29, 40, 43, 44 (See Table~\ref{tab:Mstars2MASS}) have been indicated with a star symbol in the first column of both Tables~\ref{tab:Mstars} and \ref{tab:Mstars2MASS}.

Correctly cross-identified stars should have consistent magnitudes as well as position. We computed the difference between the $J$ and $K_{\rm s}$ DENIS and 2MASS magnitudes for this set of 8 stars. The error in the magnitude difference is taken to be the sum in quadrature of the photometric errors. The DENIS and 2MASS $J$ magnitudes are consistent at the $1\sigma$ level for all stars except stars 7 and 9, which differ by $2.5\sigma$  and $3.4\sigma$ respectively. Stars  16, 29, 43 and 44 have DENIS and 2MASS $K_{\rm s}$ magnitudes consistent within $1\sigma$; stars 9, 19 and 40 differ in $K_{\rm s}$ by $1.7\sigma$, $1.3\sigma$ and $1.1\sigma$ respectively and no DENIS  $K_{\rm s}$ magnitude is available for star 7. \citet{2001AJ....121.2851C} gives relations for transforming DENIS  $K_{\rm s}$ magnitudes and $J- K_{\rm s}$ colour to the 2MASS photometry system. The transformation produces a shift of $< -0.02$ mag on the DENIS $J$ and $K_{\rm s}$ magnitudes. Including this photometry transformation results in the $<1\sigma$ consistency between DENIS and 2MASS magnitudes as above except for star 44 which has $J$ band magnitudes different by $1.1\sigma$, and stars 9, 19, 29 and 40 which differ in $K_{\rm s}$ by $2.0\sigma$,  $1.2\sigma$, $1.1\sigma$ and $1.3\sigma$ respectively. \citet{2001A&A...380..590P} however notes that the  \citet{2001AJ....121.2851C} photometry transformations were derived using a small fraction of the total DENIS data. The increased disagreement between 2MASS and DENIS magnitudes arising from the application of these transformations may therefore be not significant.  We note that blending within the seeing disks of both the  2MASS and DENIS experiments may account for small differences in photometry.

We therefore claim that stars 16, 19, 29, 40, 43 and 44 have been correctly cross-identified between the OGLE-II, DENIS and 2MASS catalogues, based on their consistent position and magnitudes which agree to within $1.3\sigma$. The entries for these 6 stars are indicated by a dagger symbol in Tables~\ref{tab:Mstars} and  \ref{tab:Mstars2MASS}.

\begin{figure}
\begin{center}
\centering\includegraphics[width=1.0\hsize]{\FigDir{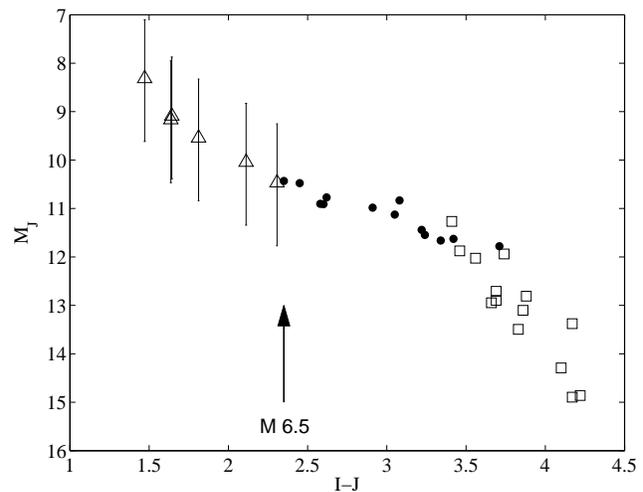}}
\end{center}
\caption{\label{fig:compWithDahn}Comparison between candidate nearby high proper motion stars in the OGLE-II proper motion catalogue with cool dwarf and brown dwarf photometry from \citet{2002AJ....124.1170D}. Triangles indicate the $J$-band absolute magnitudes, $M_{\rm J}$ and  $\ImJ$ colours for  OGLE-II high proper motion stars $(\mu > 50 \masyr)$  which have photometric distances $d<50$ pc and corresponding entries within 1 arcsec in the DENIS catalogue: stars 16, 19, 29, 40, 43 and 44 (UIDs, see Table~\ref{tab:Mstars2MASS} and text). The photometric data for late M-type stars from \citet{2002AJ....124.1170D} are shown as circles and that for stars with later spectral type are shown as squares. }
\end{figure}

The absolute $J$ band magnitude, $M_{\rm J}$, and \ImJ colour for these 6 stars is shown in Figure~\ref{fig:compWithDahn}, along with the photometry of late M-type and brown dwarf stars from \citet{2002AJ....124.1170D}. The absolute magnitude for the high proper motion stars is computed using the photometric distance estimates, $d$, and 2MASS $J$ band magnitudes in Table~\ref{tab:Mstars2MASS}. Large errors on the $M_{\rm J}$ values exist, largely arising from the uncertainly in photometric distance. \citet{2001A&A...380..590P} noted that there is a significant scatter in the late type M-dwarf colour-luminosity relation of approximately $\pm 1$ mag which corresponds to an average random error on every star of $\sim 45\%$.  \citet{2001A&A...380..590P} also find that a polynomial fit to the absolute $I$-band magnitudes as a function of $\ImJ$ colour of their sample  is up to $\sim 1$ mag lower than the theoretical luminosity model of \citet{1998A&A...337..403B}. This difference is greatest for colours $1.4 \lesssim \ImJ \lesssim 2.0$, which encompasses the majority of M-dwarf stars in our sample. If the theoretical luminosity, $M_{\rm I}$ is overestimated by 1 mag for these stars, this corresponds to a systematic $\sim 60\%$ underestimate of $d$. The lower errors on distance in Table~\ref{tab:Mstars2MASS} were computed from the sum in quadrature of the error due to DENIS $I$-band photometric uncertainties and the statistical scatter in the luminosity function. The upper distance error also includes in quadrature the systematic $60\%$ uncertainity due to the possible underestimate in $d$ arising from the  theoretical luminosity model of \citet{1998A&A...337..403B}.

The colour and magnitude of the OGLE-II high proper motion stars in Figure~\ref{fig:compWithDahn} suggests the continuation of a well-defined trend seen in the cool dwarf data from \citet{2002AJ....124.1170D}. The two bluest (and earliest) dwarf stars in \citet{2002AJ....124.1170D} have a spectral type of M6.5. All the OGLE-II nearby candidate stars are blue-ward of this, suggesting spectral types earlier than M6.5. 

\begin{figure}
\begin{center}
\centering\includegraphics[width=1.0\hsize]{\FigDir{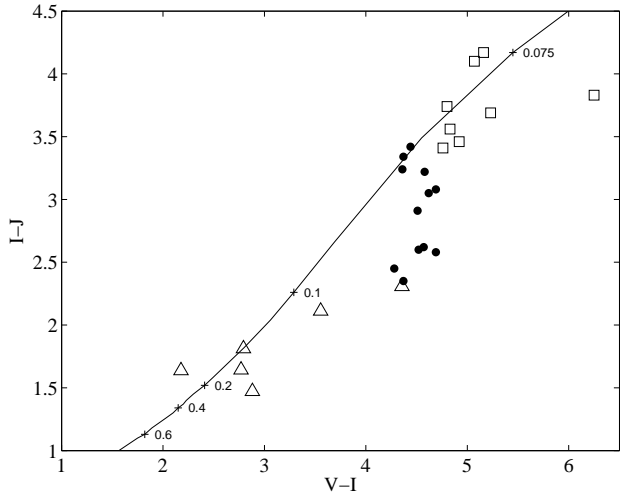}}
\end{center}
\caption{\label{fig:compWithDahnCC}\ImJ , \VmI colour-colour relation for the sample of 6 best-quality OGLE-II high proper motion stars (triangles)  with cool dwarf and brown dwarf photometry from \citet{2002AJ....124.1170D} (see caption Figure~\ref{fig:compWithDahn} for explanation of symbols). The solid black line indicates the corresponding model with solar metallicity from \citet{1998A&A...337..403B}, with stellar masses indicated in solar units.}
\end{figure}

The \ImJ,\VmI colour-colour diagram for this subsample of 6 stars is shown in Figure~\ref{fig:compWithDahnCC}. The colours of these stars are consistent with the model colours from \citet{1998A&A...337..403B}, and continues the trend observed in the \citet{2002AJ....124.1170D} data. We estimate the masses of these stars to be $0.1M_{\odot}$ - $0.3M_{\odot}$.

The star which has DENIS colour and magnitude consistent with a nearby white dwarf does not have a counterpart in the 2MASS catalogue within 5.9 arcsec. 

None of the members in the sample of candidate nearby ($d<50$ pc) high proper motion OGLE-II stars has corresponding entries in the {\it Catalogue of Nearby Stars} (CNS3, \citealt{1991adc..rept.....G}) or the catalogue of \citet{1970ROAn....5.....W}. Both catalogues have a distance limit of 25 pc. Two stars in the set of 6 best-quality candidate nearby stars have distances within 25 pc (stars 9 and 16, see Table~\ref{tab:Mstars2MASS}), however  their absence from the CNS3 catalogue suggests that this is the first distance estimate for these stars. None of these stars have corresponding entries in the {\it Hipparcos/Tycho} catalogue \citep{1997ESASP1200.....P}. However, we note that the {\it Hipparcos} limiting magnitude is $V = 12.4$ and the catalogue is complete down to $V = 7.3$ -- 9.0 mag (depending on colour) \citep{1997A&A...323L..49P}. The {\it Tycho} catalogue has a limiting magnitude of $V=11.5$ and is (99.9\%) complete to $V=10.0$ \citep{1997A&A...323L..57H}. The $V$ band magnitudes of the candidate nearby stars ($d<50$ pc) in the OGLE-II high proper motion sample are $V \gtrsim 15$, too faint for {\it Hipparcos}. 

We note that stars 13, 28 and 42 in Table~\ref{tab:Mstars2MASS} have photometric distances $d<10$ pc. This discovery of new nearby stars, if correct, is exciting. However, the closest DENIS counterparts to these OGLE-II HPM stars are at 3.01, 3.72 and 3.01 arcsecs respectively. Given the crowded nature of the Galactic bulge fields, mis-cross-identification remains a possibility that cannot be ruled out without higher resolution imaging of and around these stars.

\section{Conclusion}
\label{sec:conclusion}
The large scale microlensing survey conducted by the OGLE-II collaboration has produced proper motions for a large number of stars towards the Galactic Centre. A sample of stars with high proper motions $\mu > 50 $ \masyr was extracted and cross-identified with the DENIS infrared survey database. Stars were identified as either G/K or M dwarf stars based on their DENIS $I$-band magnitude and $\ImJ$ colour. Photometric distance estimates were derived using theoretical luminosity functions for the appropriate stellar type.

Nearby M-dwarf stars with photometric distances $d<50$ pc were selected and cross-indexed with the 2MASS catalogue. A set of 6 high proper motion stars had corresponding positions in the DENIS catalogue within 1 arcsec, cross-identified themselves with entries in the 2MASS catalogue within 1 arcsec. The DENIS and 2MASS $J$ and $K_{\rm s}$ magnitudes of these stars are consistent to within $1.3\sigma$.  The absolute $J$ band magnitude $M_{\rm J}$ and \ImJ colour of these 6 stars continues a well-defined trend seen in the sample of late type nearby M and L dwarfs given in  \citet{2002AJ....124.1170D}. Large uncertainties exist on the photometric distance estimation, arising from  the observed intrinsic scatter in the luminosity and colour of M-dwarfs and the possible systematic overestimation of these stars' theoretical absolute magnitude. However, the comparison between the \ImJ colour of these 6 stars and the data from \citet{2002AJ....124.1170D} suggest that these stars are M dwarfs with spectral type earlier than M6.5. The \ImJ , \VmI colour-colour relationship of these 6 stars clearly continues the trend seen in the \citet{2002AJ....124.1170D} data, and is consistent with theoretical model predictions from  \citet{1998A&A...337..403B}, suggesting masses for these stars approximately $0.1 M_{\odot}$ -- $0.3 M_{\odot}$.

None of the candidate nearby ($d<50$ pc) stars appear in the CNS3 or {\it Hipparcos/Tycho} catalogues.  Most proper motion surveys do not cover the Galactic plane due to the high levels of extinction encountered. Extinction has not been accounted for in this work. While the effects of extinction is expected to be minimal for the nearest stars, extinction is an additional source of uncertainty in the distance estimates for these stars.

The set of 6 nearby dwarf candidates was established by applying stringent cuts on proper motion accuracy and requiring good position and magnitude cross-identification. However, due to the crowded fields in which these stars were found, there remains a risk that these high proper motion stars have been mis-identified with stars in the DENIS and 2MASS catalogues. 

It was not possible to derive reliable trigonometric distances to these stars as their location means their astrometric parallax signature is degenerate with that of differential refraction. In order to more accurately determine the stellar type of and thereby distance to these stars, low resolution spectroscopy is required. These stars would be an ideal target for optical or IR spectroscopy.

\section*{Acknowledgments}

We thank C. {Reyl{\'e}}, W. Evans, V. Belokurov and T. Sumi for helpful discussions. NJR acknowledges financial support by a PPARC PDRA fellowship.   This work was partially supported by the European Community's Sixth Framework Marie Curie Research Training Network Programme, Contract No. MRTN-CT-2004-505183 `ANGLES'.

This publication makes use of data products from the Two Micron All Sky Survey, which is a joint project of the University of Massachusetts and the Infrared Processing and Analysis Center/California Institute of Technology, funded by the National Aeronautics and Space Administration and the National Science Foundation.

 This research has made use of the VizieR catalogue access tool, CDS, Strasbourg, France.

\onecolumn

\begin{longtable}{@{\hspace{-20pt}}c@{\hspace{-20pt}}rrrccccc}
\caption{\label{tab:GKstars}Photometric distance estimation of OGLE-II high proper motion stars with DENIS counterparts. Stars are characterized as G/K-dwarfs based on DENIS $\ImJ$ colour, and distance estimate, $d$, determined using the theoretical luminosity model of \citet{1997A&AS..125..229L}. Up to four distance estimations are given, corresponding to metallicities [Fe/H] = 0.0, -1.0, -2.0 and -2.5; however more weight should be given to the first two distance estimates, see text. Columns 1-2 are the OGLE-II field number and star ID. $\mu_{\rm l}$, $\mu_{\rm b}$ are the proper motions in Galactic longitude and latitude directions from the OGLE-II high proper motion catalogue.}\\
\hline
Field &ID\hstrut{8pt}&$\mu_{\rm l}\pm \sigma_{\mu_{\rm l}}\hstrut{6pt}$  & $\mu_{\rm b}\pm \sigma_{\mu_{\rm b}}\hstrut{6pt}$  &$I$ & $\ImJ$ & $d$  & DENIS-ID \\ 
\multicolumn{2}{c}{} & (\masyr) & (\masyr) & & & (pc) & \\
\hline
\endfirsthead
Table~\ref{tab:GKstars} continued \\
\hline
Field &ID\hstrut{8pt}&$\mu_{\rm l}\pm \sigma_{\mu_{\rm l}}\hstrut{6pt}$  & $\mu_{\rm b}\pm \sigma_{\mu_{\rm b}}\hstrut{6pt}$  &$I$ & $\ImJ$ & $d$  & DENIS-ID \\ 
\multicolumn{2}{c}{} & (\masyr) & (\masyr) & & & (pc) & \\
\hline
\endhead
\hline
\endfoot
3 & 24531 & 43.29 \ppm 4.74 & -37.43 \ppm 5.31 &  14.18 & 0.995 &  438/ - / - /209  & J175310.2-301747 \\
3 & 147285 & -72.90 \ppm 0.41 & 11.60 \ppm 0.39 &  14.36 & 0.757 &  756/ - /303/ -   & J175315.0-294440 \\
3 & 287539 & -49.31 \ppm 2.97 & -29.76 \ppm 2.68 &  10.88 & 0.713 &  159/136/ 60/ -   & J175328.1-300353 \\
3 & 383035 & -53.24 \ppm 0.45 & -50.81 \ppm 0.48 &  12.17 & 0.714 &  288/250/110/ -   & J175321.7-294013 \\
4 & 13254 & -75.46 \ppm 6.54 & 28.19 \ppm 7.54 &  11.90 & 0.602 &  288/144/ 95/ -   & J175416.5-300455 \\
5 & 222816 & -82.73 \ppm 2.00 & 15.62 \ppm 1.66 &  11.24 & 0.533 &  230/ 97/ 69/ -   & J175021.5-301249 \\
6 & 319492 & -52.71 \ppm 0.59 & 17.75 \ppm 0.68 &  12.14 & 0.700 &  287/223/108/ -   & J180814.4-320952 \\
6 & 443524 & -54.00 \ppm 0.37 & -15.87 \ppm 0.41 &  14.94 & 0.846 &  775/ - /401/268  & J180826.1-321010 \\
8 & 18351 & -56.42 \ppm 0.48 & -30.47 \ppm 0.47 &  14.11 & 0.873 &  509/ - /279/187  & J182243.1-220530 \\
8 & 57591 & -68.29 \ppm 0.57 & -35.56 \ppm 0.56 &  14.37 & 0.931 &  526/ - / - /219  & J182242.1-214402 \\
8 & 285968 & -29.89 \ppm 0.57 & -48.44 \ppm 0.62 &  12.74 & 0.567 &  442/202/138/ -   & J182312.1-213256 \\
9 & 27834 & -41.03 \ppm 0.75 & -38.52 \ppm 1.02 &  11.74 & 0.833 &  200/ - / 92/ 61  & J182342.4-215658 \\
9 & 54021 & 53.13 \ppm 4.61 & -43.33 \ppm 5.37 &  15.42 & 0.676 & 1333/807/485/ -   & J182333.4-213921 \\
9 & 279206 & -68.79 \ppm 0.82 & -8.47 \ppm 0.72 &  14.16 & 0.711 &  723/612/274/ -   & J182429.3-215312 \\
10 & 352468 & -49.91 \ppm 1.95 & 4.85 \ppm 1.62 &  12.10 & 0.889 &  197/ - /112/ 75  & J182031.3-224929 \\
13 & 25882 & -48.01 \ppm 2.24 & -25.81 \ppm 1.88 &  14.28 & 0.776 &  715/ - /293/ -   & J181640.8-241520 \\
13 & 315786 & -75.25 \ppm 4.67 & -124.77 \ppm 4.55 &  10.61 & 0.999 &   84/ - / - / 40  & J181706.2-241152 \\
13 & 563636 & -45.57 \ppm 4.70 & 62.92 \ppm 4.59 &  13.19 & 0.564 &  545/247/170/ -   & J181718.9-233259 \\
14 & 136405 & -59.59 \ppm 1.01 & 27.17 \ppm 0.99 &  13.77 & 0.906 &  414/ - /245/163  & J174633.0-224901 \\
16 & 65346 & -121.42 \ppm 0.70 & 30.17 \ppm 0.73 &  15.29 & 0.624 & 1334/708/452/ -   & J180937.7-262350 \\
16 & 501033 & 6.83 \ppm 0.86 & -69.54 \ppm 0.77 &  12.23 & 0.907 &  203/ - /120/ 80  & J181011.8-255821 \\
16 & 543398 & -125.42 \ppm 0.92 & -13.62 \ppm 0.89 &  12.13 & 0.504 &  358/144/103/ -   & J181032.7-264204 \\
18 & 51131 & -37.10 \ppm 0.53 & -47.98 \ppm 0.55 &  12.83 & 0.940 &  255/ - / - /108  & J180635.3-272354 \\
18 & 642493 & -35.86 \ppm 0.64 & -45.68 \ppm 0.68 &  12.54 & 0.829 &  291/ - /133/ 88  & J180734.4-271940 \\
19 & 73428 & -34.29 \ppm 0.68 & -48.84 \ppm 0.62 &  12.54 & 0.829 &  291/ - /133/ 88  & J180734.4-271940 \\
20 & 775068 & -55.15 \ppm 0.49 & -29.18 \ppm 0.48 &  13.73 & 0.778 &  553/ - /227/ -   & J175937.7-282547 \\
22 & 413569 & -63.30 \ppm 0.43 & -25.52 \ppm 0.46 &  12.21 & 0.801 &  264/ - /114/ 74  & J175700.7-310318 \\
24 & 77382 & -45.86 \ppm 0.63 & -20.86 \ppm 0.71 &  11.86 & 0.690 &  256/183/ 95/ -   & J175254.8-325010 \\
25 & 438478 & -53.46 \ppm 0.96 & -19.08 \ppm 1.26 &  11.88 & 0.752 &  243/209/ 97/ -   & J175438.3-323401 \\
26 & 165925 & -69.39 \ppm 0.55 & -44.44 \ppm 0.51 &  14.54 & 0.820 &  744/ - /333/220  & J174645.8-343514 \\
28 & 83963 & -169.87 \ppm 1.91 & 39.15 \ppm 1.68 &  11.51 & 0.763 &  202/ - / 81/ -   & J174641.4-364847 \\
28 & 84079 & -178.76 \ppm 6.18 & 43.54 \ppm 7.03 &  11.51 & 0.763 &  202/ - / 81/ -   & J174641.4-364847 \\
31 & 124959 & -41.53 \ppm 2.78 & -32.29 \ppm 3.88 &  13.47 & 0.701 &  530/414/199/ -   & J180158.6-282843 \\
31 & 765538 & -110.95 \ppm 0.60 & -68.70 \ppm 0.60 &  14.13 & 0.969 &  444/ - / - /200  & J180243.3-281355 \\
33 & 771 & -67.65 \ppm 0.94 & -8.80 \ppm 1.03 &  15.39 & 0.914 &  861/ - /519/345  & J180507.4-291905 \\
33 & 365983 & -58.09 \ppm 0.48 & -31.30 \ppm 0.49 &  13.53 & 0.665 &  564/333/203/ -   & J180516.9-282951 \\
36 & 648617 & -32.92 \ppm 0.49 & 43.85 \ppm 0.52 &  13.64 & 0.944 &  369/ - / - /158  & J180534.2-273148 \\
37 & 181412 & -53.67 \ppm 1.02 & 12.11 \ppm 0.98 &  11.60 & 0.739 &  216/177/ 85/ -   & J175228.1-301557 \\
38 & 45507 & -37.57 \ppm 0.74 & -34.68 \ppm 0.77 &  11.91 & 0.915 &  173/ - /105/ 70  & J180110.2-300803 \\
39 & 222196 & -74.38 \ppm 0.52 & -2.69 \ppm 0.53 &  12.12 & 0.846 &  212/ - /110/ 73  & J175533.7-301125 \\
39 & 387848 & -44.83 \ppm 3.44 & -30.74 \ppm 2.75 &  10.98 & 0.553 &  200/ 88/ 62/ -   & J175531.7-292516 \\
40 & 246654 & -67.89 \ppm 0.46 & -6.69 \ppm 0.54 &  12.85 & 0.948 &  254/ - / - /109  & J175051.7-331230 \\
40 & 452942 & -68.36 \ppm 0.52 & 25.24 \ppm 0.55 &  12.71 & 0.652 &  393/224/139/ -   & J175114.6-325536 \\
41 & 16970 & -56.46 \ppm 0.50 & -9.61 \ppm 0.49 &  14.45 & 0.831 &  701/ - /320/213  & J175139.9-332813 \\
41 & 473835 & -95.23 \ppm 0.50 & -46.73 \ppm 0.51 &  12.73 & 0.972 &  232/ - / - /105  & J175233.6-332631 \\
41 & 503739 & -68.47 \ppm 2.94 & -134.26 \ppm 2.52 &  11.57 & 0.899 &  152/ - / 88/ 59  & J175238.9-331449 \\
42 & 445360 & -57.07 \ppm 0.97 & -44.35 \ppm 0.98 &  11.81 & 0.998 &  146/ - / - / 70  & J180908.7-263004 \\
44 & 265613 & -29.96 \ppm 0.48 & -46.23 \ppm 0.52 &  12.62 & 0.606 &  399/202/132/ -   & J174945.1-300610 \\
45 & 10225 & -51.10 \ppm 0.86 & -3.74 \ppm 0.97 &  12.44 & 0.820 &  284/ - /127/ 84  & J180314.7-302919 \\
46 & 26214 & -109.74 \ppm 4.29 & 40.18 \ppm 3.24 &  11.47 & 0.886 &  148/ - / 84/ 56  & J180408.1-302031 \\
46 & 298227 & -139.69 \ppm 6.23 & 44.77 \ppm 4.70 &  12.96 & 0.734 &  405/349/158/ -   & J180455.3-303057 \\
47 & 23616 & -59.23 \ppm 1.61 & -7.29 \ppm 1.52 &  14.89 & 0.917 &  682/ - /414/275  & J172627.8-395651 \\
47 & 28899 & -50.05 \ppm 1.11 & 9.08 \ppm 1.47 &  13.58 & 0.595 &  630/310/205/ -   & J172639.2-395338 \\
47 & 141928 & -58.29 \ppm 1.26 & 38.87 \ppm 1.13 &  12.86 & 0.798 &  359/ - /153/100  & J172657.6-392324 \\
47 & 163215 & -57.10 \ppm 0.71 & 0.17 \ppm 0.72 &  15.31 & 0.774 & 1153/ - /471/ -   & J172714.5-400421 \\
49 & 152288 & -29.27 \ppm 0.82 & 40.82 \ppm 0.91 &  12.23 & 0.631 &  323/175/111/ -   & J172939.0-402629 \\
49 & 152290 & -49.12 \ppm 0.82 & -11.10 \ppm 0.87 &  12.85 & 0.972 &  246/ - / - /112  & J172940.1-402623 \\

\end{longtable}

\setlongtables
\begin{longtable}{@{\hspace{15pt}}l@{\hspace{-50pt}}rrrccccc}
\caption{\label{tab:Mstars}Photometric distance estimation of OGLE-II high proper motion stars with DENIS counterparts. Stars are characterized as M-dwarfs based on DENIS $\ImJ$ colour. The distance $d$  is determined using the theoretical luminosity model of \citet{1998A&A...337..403B} for a solar metallicity star with errors derived from the uncertainty in the reported $I$-band magnitudes. Columns 1-2 are the OGLE-II field number and star ID. $\mu_{\rm l}$, $\mu_{\rm b}$ are the proper motions in Galactic longitude and latitude directions from the OGLE-II high proper motion catalogue. See Section~\ref{sec:discussion} for an explanation of the star and dagger symbols in column 1.} \\
\hline
\hspace{-10pt}Field & ID\hstrut{5pt}  &$\mu_{\rm l}\pm \sigma_{\mu_{\rm l}}\hstrut{8pt}$  & $\mu_{\rm b}\pm \sigma_{\mu_{\rm b}}\hstrut{2pt}$  &$I$ & $\ImJ$ & $d$  & DENIS-ID \\ 
\multicolumn{2}{c}{} & (\masyr)\hstrut{2pt} & (\masyr)\hstrut{2pt} & & & (pc) & \\
\hline
\endfirsthead
Table~\ref{tab:Mstars} continued \\
\hline
\hspace{-10pt}Field & ID\hstrut{5pt}  &$\mu_{\rm l}\pm \sigma_{\mu_{\rm l}}\hstrut{2pt}$  & $\mu_{\rm b}\pm \sigma_{\mu_{\rm b}}\hstrut{2pt}$  &$I$ & $\ImJ$ & $d$  & DENIS-ID \\ 
\multicolumn{2}{c}{} & (\masyr)\hstrut{2pt} & (\masyr)\hstrut{2pt} & & & (pc) & \\
\hline
\endhead
\hline
\endfoot
1 & 202039 & -64.28 \ppm 0.58 & -29.93 \ppm 0.66 &  15.01 & 1.527 & 	$ 94^{  +3}_{  -3}$ & J180219.8-301955 \\
1 & 238184 & 45.12 \ppm 4.47 & -30.93 \ppm 6.12 &  15.16 & 1.916 & 	$ 50^{  +2}_{  -2}$ & J180222.9-301005 \\
1 & 587955 & -44.03 \ppm 3.42 & -26.51 \ppm 3.09 &  15.31 & 1.713 & 	$ 72^{  +3}_{  -3}$ & J180250.1-301340 \\
1 & 619578 & 8.66 \ppm 2.46 & -73.00 \ppm 2.74 &  15.04 & 1.609 & 	$ 78^{  +3}_{  -3}$ & J180249.0-300128 \\
1 & 701357 & -92.55 \ppm 5.20 & -9.90 \ppm 4.33 &  13.48 & 1.555 & 	$ 43^{  +1}_{  -1}$ & J180259.7-293831 \\
2 & 413124 & -72.61 \ppm 0.86 & 20.75 \ppm 0.75 &  15.11 & 1.461 & 	$123^{  +4}_{  -4}$ & J180417.4-282734 \\
2 & 426502 & -56.39 \ppm 5.29 & 17.54 \ppm 5.03 &  12.91 & 1.758 & 	$ 22^{  +1}_{  -1}$ & J180429.7-291729 \\
3 & 64760 & 2.99 \ppm 6.52 & -69.43 \ppm 6.51 &  15.09 & 1.997 & 	$ 43^{  +2}_{  -2}$ & J175301.9-300832 \\
3 & 216180 & -61.62 \ppm 5.49 & 2.65 \ppm 4.50 &  13.38 & 2.253 & 	$ 15^{  +1}_{  -1}$ & J175330.8-302535 \\
3 & 229801 & 44.99 \ppm 4.38 & -22.30 \ppm 4.80 &  13.65 & 2.056 & 	$ 21^{  +1}_{  -1}$ & J175327.9-301913 \\
3 & 301228 & -51.62 \ppm 5.06 & 9.91 \ppm 4.20 &  13.26 & 1.323 & 	$ 96^{  +2}_{  -2}$ & J175323.5-295901 \\
3$^{\star}$ & 370016 & -164.93 \ppm 0.90 & -107.34 \ppm 1.00 &  13.83 & 1.834 & 	$ 30^{  +1}_{  -1}$ & J175329.3-294140 \\
4 & 230257 & -49.96 \ppm 0.60 & -6.48 \ppm 0.71 &  13.15 & 1.020 & 	$215^{  +6}_{  -5}$ & J175426.5-300256 \\
4 & 231890 & -52.81 \ppm 4.75 & 20.29 \ppm 5.00 &  14.39 & 1.829 & 	$ 40^{  +2}_{  -2}$ & J175431.3-300158 \\
4$^{\star}$ & 254212 & -131.63 \ppm 1.99 & 72.61 \ppm 2.73 &  11.89 & 1.508 & 	$ 24^{  +1}_{  -1}$ & J175429.1-295718 \\
4 & 421737 & -105.40 \ppm 8.75 & 47.93 \ppm 9.43 &  14.90 & 1.696 & 	$ 61^{  +2}_{  -2}$ & J175441.2-300520 \\
4 & 464317 & -52.12 \ppm 5.45 & 27.58 \ppm 5.86 &  12.81 & 1.007 & 	$188^{  +4}_{  -4}$ & J175442.7-295344 \\
4 & 614523 & -72.49 \ppm 5.16 & 45.21 \ppm 6.71 &  13.43 & 1.977 & 	$ 21^{  +1}_{  -1}$ & J175454.4-300223 \\
4 & 623507 & -82.21 \ppm 0.44 & -57.50 \ppm 0.47 &  13.66 & 1.511 & 	$ 53^{  +2}_{  -2}$ & J175453.8-295925 \\
4 & 625701 & -48.45 \ppm 5.16 & 55.79 \ppm 8.22 &  14.89 & 2.038 & 	$ 37^{  +2}_{  -2}$ & J175457.0-300005 \\
4 & 734288 & -188.71 \ppm 15.85 & 17.44 \ppm 12.94 &  15.04 & 2.669 & 	$ 21^{  +1}_{  -1}$ & J175503.6-292621 \\
5 & 38999 & -38.77 \ppm 1.57 & -40.39 \ppm 1.86 &  14.00 & 3.087 & 	$  9^{  +1}_{  -1}$ & J174950.1-300004 \\
5 & 208658 & -57.71 \ppm 0.42 & 27.11 \ppm 0.51 &  14.01 & 2.652 & 	$ 13^{  +1}_{  -1}$ & J175027.5-301923 \\
5 & 244353 & -10.38 \ppm 2.80 & -134.98 \ppm 3.20 &  16.08 & 2.723 & 	$ 32^{  +2}_{  -2}$ & J175036.2-300145 \\
5 & 314114 & -55.19 \ppm 0.38 & -78.68 \ppm 0.40 &  13.81 & 1.325 & 	$123^{  +2}_{  -2}$ & J175046.2-302213 \\
5 & 372152 & 30.62 \ppm 0.52 & -101.08 \ppm 0.57 &  13.83 & 1.416 & 	$ 81^{  +2}_{  -2}$ & J175051.3-295803 \\
6 & 30011 & -57.46 \ppm 0.65 & -25.04 \ppm 0.64 &  14.96 & 1.159 & 	$357^{ +11}_{ -10}$ & J180732.1-321842 \\
6 & 351465 & -52.31 \ppm 0.52 & -39.74 \ppm 0.49 &  14.18 & 1.170 & 	$244^{  +6}_{  -6}$ & J180819.4-315412 \\
7 & 356532 & -48.59 \ppm 1.04 & -38.63 \ppm 1.04 &  16.22 & 1.175 & 	$617^{ +27}_{ -26}$ & J180942.6-323124 \\
8 & 50479 & -51.31 \ppm 0.90 & -1.10 \ppm 0.83 &  16.74 & 1.589 & 	$178^{ +10}_{  -9}$ & J182249.2-214745 \\
8$^{\star,\dagger}$ & 98776 & -12.37 \ppm 3.27 & -62.42 \ppm 4.06 &  13.80 & 2.112 & 	$ 21^{  +1}_{  -1}$ & J182247.8-212100 \\
8 & 197686 & 90.86 \ppm 0.60 & -12.55 \ppm 0.70 &  13.95 & 1.418 & 	$ 85^{  +2}_{  -2}$ & J182305.6-212328 \\
8 & 240063 & -119.22 \ppm 1.41 & -87.68 \ppm 1.87 &  12.37 & 1.280 & 	$ 75^{  +1}_{  -1}$ & J182320.7-215712 \\
8 & 350020 & -50.64 \ppm 1.73 & -31.68 \ppm 2.11 &  13.68 & 1.565 & 	$ 46^{  +1}_{  -1}$ & J182326.3-214906 \\
8 & 397026 & 24.30 \ppm 0.80 & -77.33 \ppm 0.78 &  16.18 & 1.715 & 	$107^{  +4}_{  -4}$ & J182329.4-212300 \\
9 & 5893 & -65.82 \ppm 0.54 & -12.08 \ppm 0.62 &  16.40 & 1.435 & 	$245^{ +10}_{  -9}$ & J182346.4-221211 \\
9 & 68237 & -68.36 \ppm 0.58 & -7.54 \ppm 0.56 &  15.52 & 1.563 & 	$108^{  +3}_{  -3}$ & J182336.9-212713 \\
9 & 177063 & -15.63 \ppm 0.50 & 57.80 \ppm 0.53 &  16.06 & 1.558 & 	$140^{  +5}_{  -5}$ & J182408.1-220828 \\
9 & 304725 & -53.50 \ppm 0.88 & 3.11 \ppm 1.03 &  15.32 & 1.419 & 	$158^{ +11}_{ -10}$ & J182422.8-213618 \\
10 & 7699 & -54.28 \ppm 0.93 & -7.97 \ppm 0.87 &  16.71 & 1.126 & 	$862^{ +37}_{ -36}$ & J181942.3-224617 \\
10 & 92354 & -98.88 \ppm 1.88 & -37.97 \ppm 1.82 &  17.30 & 1.745 & 	$172^{ +11}_{ -10}$ & J181944.2-220635 \\
10 & 106322 & -67.70 \ppm 0.66 & 15.29 \ppm 0.62 &  14.74 & 1.359 & 	$162^{  +4}_{  -4}$ & J181951.2-220045 \\
10 & 441733 & -92.28 \ppm 1.53 & -2.61 \ppm 1.23 &  15.52 & 1.202 & 	$416^{+244}_{-154}$ & J182032.9-220222 \\
11 & 121881 & -64.58 \ppm 2.03 & -9.95 \ppm 2.79 &  15.04 & 1.734 & 	$ 62^{  +2}_{  -2}$ & J182105.0-224434 \\
11 & 314503 & -50.26 \ppm 1.62 & 6.81 \ppm 1.53 &  15.63 & 1.616 & 	$101^{  +3}_{  -3}$ & J182112.1-215905 \\
11 & 341519 & -58.53 \ppm 0.57 & -28.50 \ppm 0.59 &  13.22 & 1.066 & 	$200^{  +3}_{  -3}$ & J182122.3-224108 \\
11 & 409511 & -10.78 \ppm 2.79 & -48.89 \ppm 3.04 &  15.81 & 1.701 & 	$ 93^{  +4}_{  -3}$ & J182125.3-220342 \\
12 & 51118 & -33.20 \ppm 1.66 & -37.58 \ppm 2.09 &  13.12 & 1.851 & 	$ 21^{  +1}_{  -1}$ & J181541.4-240408 \\
12 & 96416 & -32.53 \ppm 2.07 & -43.47 \ppm 3.07 &  14.15 & 1.456 & 	$ 81^{  +2}_{  -2}$ & J181548.0-234523 \\
12 & 206648 & -78.01 \ppm 2.79 & 30.14 \ppm 3.08 &  15.62 & 1.571 & 	$111^{  +4}_{  -4}$ & J181552.0-235439 \\
12 & 300639 & -67.88 \ppm 0.55 & 1.62 \ppm 0.53 &  13.48 & 1.113 & 	$202^{  +4}_{  -4}$ & J181610.3-241424 \\
12 & 317171 & -14.68 \ppm 0.46 & -60.23 \ppm 0.48 &  14.39 & 1.380 & 	$123^{  +3}_{  -3}$ & J181621.4-240520 \\
12 & 373469 & -168.15 \ppm 0.65 & -47.15 \ppm 0.69 &  16.19 & 1.998 & 	$ 71^{  +4}_{  -3}$ & J181617.3-234339 \\
12 & 492778 & -67.20 \ppm 1.29 & -13.68 \ppm 1.26 &  13.09 & 1.076 & 	$184^{  +4}_{  -4}$ & J181636.0-234538 \\
12 & 493160 & -70.56 \ppm 1.96 & -17.01 \ppm 1.39 &  16.33 & 1.669 & 	$125^{  +6}_{  -6}$ & J181635.1-234542 \\
13 & 98184 & -68.53 \ppm 0.71 & -13.94 \ppm 0.81 &  13.09 & 1.076 & 	$184^{  +4}_{  -4}$ & J181636.0-234538 \\
13 & 98450 & -78.63 \ppm 1.51 & -17.89 \ppm 1.06 &  16.33 & 1.669 & 	$125^{  +6}_{  -6}$ & J181635.1-234542 \\
13 & 413466 & -48.39 \ppm 0.76 & 13.41 \ppm 0.70 &  14.95 & 1.450 & 	$119^{  +3}_{  -3}$ & J181711.4-233504 \\
14 & 417830 & -52.23 \ppm 0.69 & -0.94 \ppm 0.71 &  12.10 & 1.172 & 	$ 93^{  +2}_{  -2}$ & J174703.0-225835 \\
14$^{\star,\dagger}$ & 525442 & -87.78 \ppm 0.92 & -61.73 \ppm 1.11 &  15.66 & 2.307 & 	$ 40^{  +2}_{  -2}$ & J174719.4-231412 \\
14 & 545918 & -127.05 \ppm 1.24 & 110.43 \ppm 0.96 &  15.59 & 1.462 & 	$153^{  +6}_{  -5}$ & J174721.8-230445 \\
15 & 40035 & -88.56 \ppm 2.04 & -28.55 \ppm 1.73 &  15.71 & 1.925 & 	$ 63^{  +3}_{  -3}$ & J174736.7-231945 \\
15 & 309307 & -54.96 \ppm 2.48 & 21.05 \ppm 1.86 &  15.42 & 1.522 & 	$115^{  +5}_{  -5}$ & J174806.6-224018 \\
15 & 371749 & -87.07 \ppm 1.00 & -19.86 \ppm 1.49 &  14.75 & 1.317 & 	$195^{  +9}_{  -8}$ & J174817.3-231154 \\
15 & 470245 & -71.40 \ppm 0.97 & -2.65 \ppm 1.08 &  14.05 & 1.130 & 	$251^{ +10}_{ -10}$ & J174832.7-232803 \\
15 & 521758 & -70.25 \ppm 1.63 & -0.06 \ppm 1.58 &  13.99 & 1.647 & 	$ 44^{  +2}_{  -2}$ & J174832.0-230851 \\
15 & 557293 & -16.67 \ppm 0.49 & -70.29 \ppm 0.46 &  14.16 & 1.415 & 	$ 94^{  +4}_{  -4}$ & J174825.9-225216 \\
16 & 41668 & -44.35 \ppm 0.49 & -27.79 \ppm 0.58 &  14.38 & 1.428 & 	$ 99^{  +2}_{  -2}$ & J180948.4-263009 \\
16 & 319688 & -53.08 \ppm 0.50 & -16.73 \ppm 0.49 &  14.38 & 1.425 & 	$100^{  +2}_{  -2}$ & J180959.5-260305 \\
17 & 115581 & -81.21 \ppm 0.73 & -0.63 \ppm 0.61 &  15.93 & 1.280 & 	$385^{ +85}_{ -70}$ & J181040.8-260435 \\
17 & 395098 & 46.73 \ppm 4.73 & -17.95 \ppm 5.77 &  15.06 & 1.539 & 	$ 93^{  +3}_{  -3}$ & J181112.9-262949 \\
17 & 566438 & -75.45 \ppm 0.51 & -34.81 \ppm 0.52 &  13.18 & 1.185 & 	$148^{  +3}_{  -3}$ & J181128.5-262343 \\
18 & 555084 & -30.16 \ppm 0.97 & -39.90 \ppm 0.79 &  14.63 & 1.102 & 	$351^{ +10}_{ -10}$ & J180712.2-265031 \\
19 & 202597 & -172.97 \ppm 1.02 & -1.00 \ppm 1.21 &  15.52 & 1.588 & 	$102^{  +3}_{  -3}$ & J180752.3-273740 \\
19 & 236380 & -56.14 \ppm 2.02 & -2.62 \ppm 1.92 &  15.53 & 1.673 & 	$ 86^{  +3}_{  -3}$ & J180800.3-272746 \\
19 & 258847 & -55.23 \ppm 0.44 & -14.93 \ppm 0.45 &  14.56 & 2.219 & 	$ 26^{  +1}_{  -1}$ & J180753.1-272126 \\
19 & 450718 & -54.86 \ppm 4.78 & 16.80 \ppm 7.18 &  13.74 & 1.442 & 	$ 70^{  +2}_{  -2}$ & J180803.8-271759 \\
19 & 470116 & -57.04 \ppm 0.47 & 26.99 \ppm 0.44 &  14.68 & 1.022 & 	$432^{ +11}_{ -10}$ & J180805.9-271202 \\
20 & 445532 & 49.24 \ppm 4.67 & -32.80 \ppm 5.98 &  13.53 & 1.775 & 	$ 29^{  +1}_{  -1}$ & J175934.3-291030 \\
20 & 596512 & -109.77 \ppm 1.22 & -63.35 \ppm 1.45 &  13.68 & 2.533 & 	$ 13^{  +1}_{  -1}$ & J175949.8-291926 \\
20 & 625107 & -307.64 \ppm 28.69 & -127.48 \ppm 17.38 &  14.65 & 1.572 & 	$ 71^{  +2}_{  -2}$ & J175935.6-291157 \\
20 & 632206 & -71.79 \ppm 1.51 & -23.48 \ppm 1.63 &  13.05 & 1.142 & 	$154^{  +3}_{  -3}$ & J175948.0-290733 \\
20 & 764809 & -54.03 \ppm 3.93 & 7.63 \ppm 4.69 &  15.32 & 1.660 & 	$ 80^{  +3}_{  -3}$ & J175949.1-283047 \\
21 & 140 & -110.20 \ppm 1.40 & -65.89 \ppm 1.71 &  13.68 & 2.533 & 	$ 13^{  +1}_{  -1}$ & J175949.8-291926 \\
21 & 254557 & -45.24 \ppm 3.00 & 31.95 \ppm 3.00 &  15.37 & 1.644 & 	$ 84^{  +3}_{  -3}$ & J180021.9-291515 \\
21 & 298351 & -212.48 \ppm 25.65 & -155.58 \ppm 18.58 &  15.11 & 1.271 & 	$274^{  +9}_{  -9}$ & J180012.0-290346 \\
21 & 323807 & -46.00 \ppm 0.43 & -36.19 \ppm 0.46 &  14.40 & 1.277 & 	$193^{  +5}_{  -5}$ & J180017.0-285728 \\
21 & 379680 & -43.24 \ppm 0.54 & 41.45 \ppm 0.56 &  13.27 & 1.033 & 	$221^{  +5}_{  -5}$ & J180022.0-284122 \\
21 & 422978 & -60.61 \ppm 1.92 & 26.60 \ppm 2.18 &  14.52 & 1.184 & 	$275^{  +7}_{  -7}$ & J180016.3-283333 \\
21 & 480442 & 23.64 \ppm 3.66 & 50.41 \ppm 4.50 &  17.05 & 1.947 & 	$114^{  +8}_{  -7}$ & J180022.2-291607 \\
21 & 649329 & -114.46 \ppm 0.48 & -17.12 \ppm 0.50 &  13.41 & 1.620 & 	$ 36^{  +1}_{  -1}$ & J180023.9-282744 \\
22 & 155458 & -70.72 \ppm 0.98 & 35.92 \ppm 1.45 &  16.14 & 2.204 & 	$ 55^{  +2}_{  -2}$ & J175620.1-303231 \\
22 & 681768 & -40.85 \ppm 1.15 & 29.50 \ppm 1.44 &  15.25 & 1.983 & 	$ 47^{  +2}_{  -2}$ & J175705.2-303146 \\
23 & 63756 & -73.23 \ppm 1.77 & -4.45 \ppm 1.68 &  13.63 & 1.518 & 	$ 51^{  +1}_{  -1}$ & J175726.3-312132 \\
23 & 281219 & -33.18 \ppm 2.59 & -63.29 \ppm 2.77 &  14.74 & 1.797 & 	$ 49^{  +2}_{  -2}$ & J175740.8-311443 \\
23 & 339709 & -16.76 \ppm 0.53 & -109.97 \ppm 0.50 &  15.96 & 1.663 & 	$107^{  +5}_{  -5}$ & J175748.2-305717 \\
23 & 534675 & -58.43 \ppm 2.63 & -23.89 \ppm 1.86 &  16.12 & 1.917 & 	$ 77^{  +4}_{  -4}$ & J175757.6-304930 \\
24 & 201442 & -147.36 \ppm 13.76 & 8.91 \ppm 11.46 &  16.21 & 1.708 & 	$110^{  +5}_{  -5}$ & J175311.8-330610 \\
24 & 266528 & -114.14 \ppm 4.41 & 16.47 \ppm 3.05 &  15.62 & 1.715 & 	$ 83^{  +3}_{  -3}$ & J175314.9-324129 \\
24 & 334946 & -55.86 \ppm 0.40 & -31.26 \ppm 0.47 &  12.63 & 1.276 & 	$ 86^{  +2}_{  -2}$ & J175331.8-331227 \\
25 & 327263 & -56.25 \ppm 0.39 & 12.27 \ppm 0.43 &  13.14 & 3.304 & 	$  5^{  +1}_{  -1}$ & J175441.4-331626 \\
25$^{\star,\dagger}$ & 592433 & -158.28 \ppm 0.77 & -35.01 \ppm 0.98 &  12.05 & 1.472 & 	$ 29^{  +1}_{  -1}$ & J175443.1-323213 \\
26 & 165846 & -54.27 \ppm 0.85 & 7.91 \ppm 0.93 &  14.10 & 1.085 & 	$286^{  +4}_{  -4}$ & J174641.8-343514 \\
26 & 275718 & -106.83 \ppm 3.15 & -66.45 \ppm 2.71 &  13.54 & 1.354 & 	$ 95^{  +2}_{  -2}$ & J174700.9-345618 \\
26 & 340032 & -75.97 \ppm 1.00 & 21.38 \ppm 1.30 &  15.28 & 1.244 & 	$327^{  +8}_{  -8}$ & J174706.4-343934 \\
26 & 585326 & -47.81 \ppm 4.78 & 38.55 \ppm 6.74 &  15.98 & 1.067 & 	$713^{ +24}_{ -23}$ & J174739.7-351428 \\
26 & 604525 & -65.52 \ppm 6.45 & -3.94 \ppm 5.40 &  15.23 & 1.157 & 	$407^{ +10}_{ -10}$ & J174743.7-350905 \\
27 & 152097 & -61.08 \ppm 0.54 & 3.88 \ppm 0.57 &  15.09 & 1.417 & 	$143^{  +3}_{  -3}$ & J174756.9-345017 \\
28 & 77323 & -52.28 \ppm 0.53 & -19.20 \ppm 0.53 &  16.23 & 1.497 & 	$181^{  +6}_{  -6}$ & J174636.7-365044 \\
28 & 229742 & -58.83 \ppm 0.45 & -46.38 \ppm 0.48 &  13.49 & 1.262 & 	$134^{  +2}_{  -2}$ & J174711.4-372123 \\
28 & 285427 & -109.54 \ppm 0.46 & 3.11 \ppm 0.52 &  15.03 & 1.294 & 	$241^{  +5}_{  -5}$ & J174714.8-364840 \\
28 & 328487 & -62.11 \ppm 1.66 & -2.51 \ppm 1.52 &  16.13 & 1.172 & 	$596^{ +23}_{ -22}$ & J174724.7-371933 \\
29 & 7843 & -2.59 \ppm 0.38 & -94.98 \ppm 0.46 &  15.67 & 1.704 & 	$ 87^{  +3}_{  -3}$ & J174743.7-372955 \\
29 & 39198 & -44.61 \ppm 0.73 & -22.85 \ppm 0.61 &  15.53 & 1.308 & 	$287^{  +9}_{  -8}$ & J174751.6-371522 \\
29 & 39202 & -292.94 \ppm 0.73 & 21.67 \ppm 0.68 &  16.13 & 1.056 & 	$782^{ +30}_{ -29}$ & J174750.2-371516 \\
29 & 113783 & -14.14 \ppm 1.01 & -49.68 \ppm 1.31 &  16.75 & 1.448 & 	$274^{ +15}_{ -14}$ & J174742.3-364339 \\
29 & 131780 & -51.76 \ppm 0.46 & -20.30 \ppm 0.48 &  17.23 & 1.731 & 	$170^{ +12}_{ -11}$ & J174803.3-373208 \\
29 & 160643 & -56.88 \ppm 1.41 & -14.79 \ppm 1.07 &  16.93 & 1.459 & 	$287^{ +17}_{ -16}$ & J174753.7-372121 \\
29 & 168329 & -79.27 \ppm 0.48 & -30.48 \ppm 0.53 &  13.36 & 1.031 & 	$231^{  +4}_{  -4}$ & J174801.4-371626 \\
29 & 224715 & -55.58 \ppm 0.73 & 28.91 \ppm 0.70 &  17.04 & 1.223 & 	$789^{ +45}_{ -43}$ & J174804.4-365036 \\
30 & 152246 & -40.60 \ppm 0.78 & -32.51 \ppm 0.86 &  12.00 & 1.512 & 	$ 25^{  +1}_{  -1}$ & J180107.1-283547 \\
30 & 583187 & -102.79 \ppm 1.82 & -33.05 \ppm 1.79 &  14.32 & 1.538 & 	$ 66^{  +2}_{  -2}$ & J180149.2-291502 \\
31 & 340708 & -60.31 \ppm 0.69 & -29.90 \ppm 0.86 &  13.49 & 1.395 & 	$ 75^{  +2}_{  -2}$ & J180208.7-282637 \\
31 & 562353 & -55.37 \ppm 1.99 & -28.31 \ppm 2.42 &  13.42 & 1.460 & 	$ 57^{  +1}_{  -1}$ & J180233.8-281815 \\
31 & 738841 & -72.71 \ppm 1.46 & -3.25 \ppm 1.38 &  13.84 & 1.473 & 	$ 66^{  +2}_{  -2}$ & J180252.7-282247 \\
32 & 388121 & -197.37 \ppm 6.71 & -21.23 \ppm 6.83 &  14.68 & 1.225 & 	$264^{  +7}_{  -7}$ & J180311.7-281322 \\
32 & 397369 & -186.58 \ppm 9.32 & -15.52 \ppm 9.63 &  14.68 & 1.225 & 	$264^{  +7}_{  -7}$ & J180311.7-281322 \\
32 & 555529 & -38.92 \ppm 0.47 & -66.83 \ppm 0.50 &  14.81 & 1.303 & 	$211^{  +5}_{  -5}$ & J180335.2-282237 \\
32 & 661943 & -85.71 \ppm 0.53 & 11.43 \ppm 0.68 &  15.41 & 1.436 & 	$155^{  +5}_{  -5}$ & J180348.7-284517 \\
32 & 737206 & -168.55 \ppm 0.96 & -25.28 \ppm 0.86 &  15.34 & 1.549 & 	$103^{  +3}_{  -3}$ & J180342.9-282535 \\
33 & 12568 & -86.55 \ppm 2.83 & -24.00 \ppm 3.42 &  13.94 & 1.284 & 	$152^{  +3}_{  -3}$ & J180502.1-291644 \\
33 & 16638 & -74.64 \ppm 5.78 & -9.61 \ppm 7.36 &  13.94 & 1.284 & 	$152^{  +3}_{  -3}$ & J180502.1-291644 \\
33 & 84101 & -71.53 \ppm 4.21 & -29.71 \ppm 4.06 &  14.56 & 1.496 & 	$ 84^{  +2}_{  -2}$ & J180514.6-285300 \\
33 & 566200 & -36.89 \ppm 1.47 & -39.76 \ppm 1.57 &  14.59 & 1.491 & 	$ 87^{  +3}_{  -2}$ & J180603.1-291758 \\
34 & 159315 & -62.24 \ppm 2.77 & 17.44 \ppm 2.77 &  13.96 & 1.970 & 	$ 27^{  +1}_{  -1}$ & J175752.2-285845 \\
34 & 465798 & 2.54 \ppm 0.60 & -53.74 \ppm 0.69 &  12.40 & 1.457 & 	$ 36^{  +1}_{  -1}$ & J175803.7-284425 \\
35 & 55762 & -70.98 \ppm 4.59 & 7.69 \ppm 5.65 &  15.12 & 1.466 & 	$122^{  +4}_{  -4}$ & J180357.0-280925 \\
35 & 346292 & -63.16 \ppm 0.85 & -17.71 \ppm 1.03 &  13.87 & 2.592 & 	$ 13^{  +1}_{  -1}$ & J180417.9-274610 \\
35 & 415867 & -51.94 \ppm 0.76 & -19.60 \ppm 0.74 &  14.69 & 1.568 & 	$ 73^{  +2}_{  -2}$ & J180435.1-281856 \\
35 & 692986 & -48.32 \ppm 0.60 & -28.52 \ppm 0.59 &  14.86 & 1.357 & 	$173^{  +5}_{  -4}$ & J180454.4-275157 \\
36 & 301724 & -55.66 \ppm 0.47 & 4.65 \ppm 0.49 &  15.44 & 1.431 & 	$160^{  +5}_{  -5}$ & J180524.1-280656 \\
36 & 375399 & -67.05 \ppm 4.13 & -59.15 \ppm 5.08 &  14.17 & 1.395 & 	$103^{  +2}_{  -2}$ & J180526.4-274741 \\
36 & 426923 & -64.22 \ppm 0.83 & -8.93 \ppm 0.67 &  14.32 & 1.301 & 	$169^{  +4}_{  -4}$ & J180520.4-273216 \\
36 & 454051 & -49.36 \ppm 0.89 & 26.61 \ppm 0.93 &  14.80 & 1.615 & 	$ 69^{  +2}_{  -2}$ & J180540.6-282143 \\
36 & 752090 & -52.69 \ppm 3.10 & -38.95 \ppm 2.48 &  14.19 & 1.681 & 	$ 46^{  +1}_{  -1}$ & J180547.5-275834 \\
36 & 808646 & -50.57 \ppm 2.00 & -13.22 \ppm 1.99 &  14.68 & 1.015 & 	$439^{ +11}_{ -10}$ & J180548.1-274431 \\
37 & 20539 & -35.38 \ppm 3.99 & -66.97 \ppm 3.76 &  15.63 & 1.740 & 	$ 80^{  +5}_{  -4}$ & J175202.3-301714 \\
37 & 213626 & -61.06 \ppm 0.46 & -42.74 \ppm 0.55 &  14.36 & 1.359 & 	$136^{  +5}_{  -5}$ & J175223.7-300659 \\
37 & 341121 & -34.01 \ppm 2.31 & -42.87 \ppm 2.35 &  14.06 & 1.898 & 	$ 31^{  +1}_{  -1}$ & J175233.7-302209 \\
37 & 441811 & -65.04 \ppm 0.42 & -4.30 \ppm 0.42 &  13.65 & 1.711 & 	$ 34^{  +1}_{  -1}$ & J175239.7-294552 \\
37 & 514546 & 50.09 \ppm 5.04 & -21.87 \ppm 5.98 &  14.25 & 1.974 & 	$ 30^{  +1}_{  -1}$ & J175249.8-301735 \\
38 & 190202 & -90.13 \ppm 0.55 & -0.42 \ppm 0.55 &  13.77 & 1.864 & 	$ 28^{  +1}_{  -1}$ & J180120.0-302352 \\
38 & 337622 & -72.87 \ppm 0.50 & -42.35 \ppm 0.49 &  13.21 & 1.127 & 	$171^{  +3}_{  -3}$ & J180112.8-293900 \\
39 & 309705 & 119.30 \ppm 4.49 & 82.80 \ppm 2.73 &  14.87 & 1.973 & 	$ 40^{  +2}_{  -2}$ & J175531.5-294638 \\
39 & 438588 & -117.61 \ppm 10.98 & 12.62 \ppm 9.67 &  15.22 & 1.451 & 	$134^{  +4}_{  -4}$ & J175542.4-300827 \\
39 & 622098 & -80.43 \ppm 0.49 & -12.58 \ppm 0.55 &  13.35 & 1.150 & 	$174^{  +3}_{  -3}$ & J175607.9-300852 \\
40 & 246965 & -68.37 \ppm 0.57 & -3.34 \ppm 0.74 &  15.42 & 1.269 & 	$318^{  +9}_{  -9}$ & J175051.4-331222 \\
40 & 258617 & -46.64 \ppm 1.10 & 45.75 \ppm 1.40 &  15.86 & 1.615 & 	$112^{  +4}_{  -4}$ & J175052.6-330943 \\
40 & 353609 & -47.41 \ppm 0.59 & -32.71 \ppm 0.84 &  15.07 & 1.511 & 	$101^{  +3}_{  -3}$ & J175110.4-333143 \\
41$^{\star,\dagger}$ & 47647 & -63.59 \ppm 5.68 & -7.23 \ppm 4.80 &  13.83 & 1.812 & 	$ 31^{  +1}_{  -1}$ & J175149.2-331543 \\
41 & 353119 & -47.07 \ppm 0.39 & -34.91 \ppm 0.41 &  14.06 & 2.091 & 	$ 24^{  +1}_{  -1}$ & J175221.2-331444 \\
41 & 353277 & -63.24 \ppm 0.72 & -4.09 \ppm 0.96 &  16.30 & 1.518 & 	$175^{ +11}_{ -11}$ & J175211.8-331614 \\
42 & 104424 & 56.34 \ppm 0.50 & -29.94 \ppm 0.54 &  14.04 & 1.190 & 	$218^{  +4}_{  -3}$ & J180843.4-264150 \\
42 & 133638 & -507.56 \ppm 1.59 & 18.36 \ppm 2.27 &  11.71 & 1.294 & 	$ 52^{  +1}_{  -1}$ & J180847.5-263239 \\
42 & 365611 & -60.24 \ppm 0.48 & 25.42 \ppm 0.48 &  13.83 & 1.275 & 	$150^{  +3}_{  -3}$ & J180918.2-270123 \\
42 & 401145 & -46.56 \ppm 0.85 & -58.55 \ppm 0.71 &  15.50 & 1.621 & 	$ 94^{  +3}_{  -3}$ & J180906.5-264621 \\
42 & 551215 & -18.23 \ppm 0.71 & 54.02 \ppm 0.81 &  14.24 & 1.487 & 	$ 75^{  +2}_{  -2}$ & J180924.0-264338 \\
42 & 569217 & -102.07 \ppm 2.93 & -31.90 \ppm 2.35 &  13.69 & 1.201 & 	$179^{  +3}_{  -3}$ & J180930.6-263632 \\
43 & 116810 & -142.32 \ppm 0.44 & 33.05 \ppm 0.47 &  15.08 & 1.394 & 	$158^{  +3}_{  -3}$ & J173458.9-273835 \\
43 & 216905 & -33.48 \ppm 0.90 & -45.37 \ppm 1.09 &  15.41 & 1.915 & 	$ 56^{  +2}_{  -2}$ & J173500.2-264956 \\
43 & 217063 & -63.00 \ppm 0.53 & 25.12 \ppm 0.51 &  15.76 & 1.743 & 	$ 85^{  +3}_{  -3}$ & J173500.4-265038 \\
43 & 375807 & -121.31 \ppm 6.73 & -51.42 \ppm 4.90 &  15.29 & 1.925 & 	$ 52^{  +2}_{  -2}$ & J173539.6-272808 \\
44 & 270634 & -46.52 \ppm 1.46 & -37.97 \ppm 1.70 &  14.00 & 3.087 & 	$  9^{  +1}_{  -1}$ & J174950.1-300004 \\
45 & 58948 & -52.06 \ppm 2.31 & -20.00 \ppm 2.18 &  13.74 & 1.331 & 	$116^{  +3}_{  -3}$ & J180312.2-301131 \\
45 & 189813 & -167.36 \ppm 2.54 & 83.69 \ppm 2.02 &  15.19 & 1.516 & 	$105^{  +4}_{  -4}$ & J180327.4-302522 \\
45 & 217690 & -51.35 \ppm 0.80 & -4.72 \ppm 0.98 &  14.69 & 1.356 & 	$160^{  +5}_{  -5}$ & J180320.2-301439 \\
45 & 273351 & 307.88 \ppm 21.56 & -449.27 \ppm 35.34 &  15.04 & 1.575 & 	$ 84^{  +3}_{  -3}$ & J180335.1-295510 \\
45 & 313160 & -110.41 \ppm 1.10 & -20.94 \ppm 1.33 &  14.02 & 1.374 & 	$108^{  +2}_{  -2}$ & J180321.8-294338 \\
46$^{\star,\dagger}$ & 44548 & -234.03 \ppm 1.40 & -65.68 \ppm 1.78 &  14.13 & 1.643 & 	$ 48^{  +1}_{  -1}$ & J180408.7-301305 \\
46 & 188284 & -114.48 \ppm 1.25 & -119.76 \ppm 1.19 &  12.97 & 1.355 & 	$ 73^{  +2}_{  -2}$ & J180434.7-301849 \\
46$^{\star,\dagger}$ & 335077 & -120.15 \ppm 1.48 & -5.47 \ppm 1.83 &  13.63 & 1.637 & 	$ 38^{  +1}_{  -1}$ & J180454.0-301436 \\
46 & 366546 & -81.25 \ppm 2.00 & -31.76 \ppm 2.02 &  14.01 & 1.362 & 	$114^{  +3}_{  -3}$ & J180445.7-295855 \\
46 & 419860 & -74.00 \ppm 8.92 & -110.46 \ppm 12.58 &  14.46 & 1.421 & 	$106^{  +2}_{  -2}$ & J180447.5-293836 \\
47 & 109894 & -53.25 \ppm 1.21 & -58.63 \ppm 1.09 &  15.11 & 1.554 & 	$ 92^{  +3}_{  -3}$ & J172649.5-394908 \\
47 & 110343 & -65.55 \ppm 3.37 & -70.84 \ppm 3.21 &  15.11 & 1.554 & 	$ 92^{  +3}_{  -3}$ & J172649.5-394908 \\
47 & 137581 & -64.92 \ppm 0.80 & -19.78 \ppm 0.94 &  14.03 & 1.565 & 	$ 54^{  +1}_{  -1}$ & J172652.1-392620 \\
47 & 167716 & -91.09 \ppm 0.90 & -29.39 \ppm 1.00 &  12.27 & 1.154 & 	$105^{  +1}_{  -1}$ & J172717.7-395915 \\
49 & 70067 & -62.38 \ppm 1.63 & 39.77 \ppm 1.69 &  15.70 & 1.777 & 	$ 78^{ +46}_{ -29}$ & J172910.4-404030 \\
49 & 78098 & -52.81 \ppm 1.70 & -3.73 \ppm 2.11 &  13.50 & 1.237 & 	$148^{ +87}_{ -55}$ & J172914.3-403248 \\
49 & 109397 & -97.14 \ppm 0.80 & -70.38 \ppm 0.85 &  13.75 & 1.147 & 	$211^{+124}_{ -78}$ & J172919.6-400855 \\
49 & 113409 & -53.78 \ppm 0.78 & -60.41 \ppm 0.88 &  14.73 & 1.307 & 	$200^{+117}_{ -74}$ & J172916.7-400451 \\
49 & 121079 & -129.88 \ppm 0.90 & -75.05 \ppm 1.02 &  13.71 & 1.400 & 	$ 81^{ +48}_{ -31}$ & J172917.6-395552 \\
49 & 147926 & -117.40 \ppm 1.60 & -22.21 \ppm 1.69 &  14.96 & 1.527 & 	$ 92^{ +54}_{ -34}$ & J172943.1-402853 \\
49 & 161249 & -107.87 \ppm 0.73 & -36.10 \ppm 0.76 &  14.47 & 1.354 & 	$146^{ +86}_{ -54}$ & J172932.2-401929 \\
49 & 177184 & -37.24 \ppm 1.14 & -63.34 \ppm 1.16 &  17.25 & 1.274 & 	$725^{+425}_{-268}$ & J172934.8-400434 \\
49 & 180561 & -55.53 \ppm 0.78 & -10.61 \ppm 0.90 &  16.36 & 1.627 & 	$138^{ +81}_{ -51}$ & J172934.5-400033 \\

\end{longtable}

\begin{sidewaystable}
\caption{Nearby ($d<50$ pc) OGLE-II high proper motion stars cross-identified with entries in the 2MASS catalogue. Stars are characterized as M-dwarfs based on DENIS $\ImJ$ colour. The distance estimate, $d$, is determined using the theoretical luminosity model of \citet{1998A&A...337..403B}. Error values are derived from random photometric errors, statistical colour-luminosity scatter and systematic offset of the theoretical luminosity model, see text. $\Delta\theta$  and $\Delta\phi$ are the differences in star position between the OGLE-II and DENIS catalogues  and the DENIS and  2MASS catalogues respectively. See Section~\ref{sec:discussion} for an explanation of the star and dagger symbols in column 1.}

\begin{tabular}{l@{\hspace{0pt}}c@{\hspace{0pt}}rcccccccccc}
\hline
 & \multicolumn{2}{c}{OGLE-II\hstrut{5pt}}  &  DENIS & 2MASS & $\Delta\theta$ & $\Delta\phi$ & \multicolumn{3}{c}{DENIS photometry}  & \multicolumn{2}{c}{2MASS photometry} &  $d$ \\ 
\hspace{-10pt}UID & \hspace{5pt}Field & ID\hstrut{5pt}  & ID & ID  & (as) & (as) & $I$ & $J$ & $K_{\rm s}$& $J$  & $K_{\rm s}$ & (pc)  \\
\hline
 1 & 1 & 238184  & J180222.9-301005         & 18022356-3010044  & 3.48 &  8.031 &  15.163 &  13.247 \ppm 0.13 &     &   7.679 \ppm 0.02 &   6.053 \ppm 0.03 & $ 50^{+  38}_{-  22}$ \\ 
 2 & 1 & 701357  & J180259.7-293831         & 18025972-2938310  & 3.04 &  0.554 &  13.480 &  11.925 \ppm 0.25 &  10.803 \ppm 0.15 &  11.604 \ppm 0.02 &  10.486 \ppm 0.02 & $ 43^{+  33}_{-  19}$ \\ 
 3 & 2 & 426502  & J180429.7-291729         & 18042976-2917298  & 3.35 &  0.403 &  12.915 &  11.157 \ppm 0.08 &   9.927 \ppm 0.11 &  11.184 \ppm 0.03 &   9.951 \ppm 0.03 & $ 22^{+  17}_{-  10}$ \\ 
 4 & 3 & 64760  & J175301.9-300832         & 17530194-3008328  & 3.65 &   0.327 &  15.087 &  13.090 \ppm 0.08 &     &  13.441 \ppm 0.07 &  12.272 \ppm 0.06 & $ 43^{+  33}_{-  19}$ \\ 
 5 & 3 & 216180  & J175330.8-302535         & 17533087-3025354  & 3.22 &   0.347 &  13.383 &  11.130 \ppm 0.06 &   9.474 \ppm 0.07 &  11.060 \ppm 0.03 &   9.538 \ppm 0.03 & $ 15^{+  11}_{-   6}$ \\ 
 6 & 3 & 229801  & J175327.9-301913         & 17532796-3019138  & 3.18 &   0.422 &  13.653 &  11.597 \ppm 0.06 &  10.260 \ppm 0.07 &  11.640 \ppm 0.02 &  10.284 \ppm 0.03 & $ 21^{+  16}_{-   9}$ \\ 
 7$^{\star}$ & 3 & 370016  & J175329.3-294140         & 17532930-2941398  & 0.78 &   0.887 &  13.830 &  11.996 \ppm 0.07 &     &  11.822 \ppm   &  11.271 \ppm 0.09 & $ 30^{+  23}_{-  13}$ \\ 
 8 & 4 & 231890  & J175431.3-300158         & 17543132-3001585  & 2.10 &   0.313 &  14.393 &  12.564 \ppm 0.08 &  11.378 \ppm 0.08 &  12.776 \ppm 0.06 &  11.431 \ppm 0.04 & $ 40^{+  30}_{-  17}$ \\ 
 9$^{\star}$ & 4 & 254212  & J175429.1-295718         & 17542912-2957184  & 0.65 &   0.342 &  11.887 &  10.379 \ppm 0.06 &   9.377 \ppm 0.05 &  10.624 \ppm 0.04 &   9.479 \ppm 0.03 & $ 24^{+  18}_{-  10}$ \\ 
10 & 4 & 614523  & J175454.4-300223         & 17545444-3002239  & 2.66 &   0.445 &  13.426 &  11.449 \ppm 0.06 &  10.164 \ppm 0.07 &  11.494 \ppm 0.04 &  10.186 \ppm 0.04 & $ 21^{+  16}_{-   9}$ \\ 
11 & 4 & 625701  & J175457.0-300005         & 17545702-3000061  & 3.81 &   0.762 &  14.892 &  12.854 \ppm 0.09 &  11.729 \ppm 0.09 &  12.865 \ppm   &  11.537 \ppm   & $ 37^{+  28}_{-  16}$ \\ 
12 & 4 & 734288  & J175503.6-292621         & 17550366-2926218  & 1.43 &   0.362 &  15.038 &  12.369 \ppm 0.07 &  11.060 \ppm 0.09 &  12.391 \ppm 0.05 &  11.038 \ppm 0.05 & $ 21^{+  16}_{-   9}$ \\ 
13 & 5 & 38999  & J174950.1-300004         & 17495008-3000044  & 3.01 &   0.887 &  14.002 &  10.915 \ppm 0.07 &   8.811 \ppm 0.06 &  10.792 \ppm 0.03 &   8.693 \ppm 0.03 & $  9^{+   7}_{-   4}$ \\ 
14 & 5 & 208658  & J175027.5-301923         & 17502760-3019239  & 2.66 &   0.822 &  14.014 &  11.362 \ppm 0.08 &   9.558 \ppm 0.08 &  11.356 \ppm 0.04 &   9.661 \ppm 0.03 & $ 13^{+  11}_{-   6}$ \\ 
15 & 5 & 244353  & J175036.2-300145         & 17503622-3001454  & 1.95 &   0.286 &  16.076 &  13.353 \ppm 0.11 &     &  13.427 \ppm 0.08 &  12.044 \ppm 0.06 & $ 32^{+  25}_{-  14}$ \\ 
16$^{\star,\dagger}$ & 8 & 98776  & J182247.8-212100         & 18224781-2121005  & 0.40 &   0.480 &  13.796 &  11.684 \ppm 0.09 &  10.342 \ppm 0.08 &  11.612 \ppm 0.03 &  10.283 \ppm 0.03 & $ 21^{+  16}_{-   9}$ \\ 
17 & 8 & 350020  & J182326.3-214906         & 18232632-2149060  & 2.54 &   0.512 &  13.677 &  12.112 \ppm 0.08 &  11.064 \ppm 0.07 &  12.167 \ppm 0.02 &  11.114 \ppm 0.02 & $ 46^{+  35}_{-  20}$ \\ 
18 & 12 & 51118  & J181541.4-240408         & 18154140-2404093  & 3.33 &  1.051 &  13.115 &  11.264 \ppm 0.07 &   9.923 \ppm 0.06 &  11.319 \ppm 0.02 &   9.962 \ppm 0.03 & $ 21^{+  17}_{-   9}$ \\ 
19$^{\star,\dagger}$ & 14 & 525442  & J174719.4-231412         & 17471940-2314118  & 0.79 &   0.881 &  15.656 &  13.349 \ppm 0.10 &  12.633 \ppm 0.14 &  13.451 \ppm 0.07 &  12.438 \ppm 0.06 & $ 40^{+  30}_{-  17}$ \\ 
20 & 15 & 521758  & J174832.0-230851         & 17483208-2308508  & 3.36 &   0.757 &  13.986 &  12.339 \ppm 0.08 &     &  12.428 \ppm 0.06 &  11.344 \ppm 0.06 & $ 44^{+  34}_{-  20}$ \\ 
21 & 19 & 258847  & J180753.1-272126         & 18075311-2721268  & 3.76 &   0.580 &  14.560 &  12.341 \ppm 0.11 &  10.919 \ppm 0.08 &  12.344 \ppm 0.03 &  10.891 \ppm 0.02 & $ 26^{+  20}_{-  11}$ \\ 
22 & 20 & 445532  & J175934.3-291030         & 17593434-2910305  & 2.06 &    0.038 &  13.525 &  11.750 \ppm 0.10 &  10.537 \ppm 0.09 &  11.764 \ppm 0.03 &  10.534 \ppm 0.03 & $ 29^{+  22}_{-  12}$ \\ 
23 & 20 & 596512  & J175949.8-291926         & 17594987-2919257  & 3.55 &   0.800 &  13.676 &  11.143 \ppm 0.09 &   9.806 \ppm 0.09 &  11.224 \ppm 0.04 &   9.725 \ppm 0.03 & $ 13^{+  10}_{-   5}$ \\ 
24 & 21 & 140  & J175949.8-291926         & 17594987-2919257  & 3.53 &   0.800 &  13.676 &  11.143 \ppm 0.09 &   9.806 \ppm 0.09 &  11.224 \ppm 0.04 &   9.725 \ppm 0.03 & $ 13^{+  10}_{-   5}$ \\ 
25 & 21 & 649329  & J180023.9-282744         & 18002390-2827450  & 3.04 &   0.860 &  13.414 &  11.794 \ppm 0.10 &  10.457 \ppm 0.09 &  11.797 \ppm 0.04 &  10.551 \ppm 0.04 & $ 36^{+  28}_{-  16}$ \\ 
26 & 22 & 681768  & J175705.2-303146         & 17570526-3031455  & 3.33 &   0.949 &  15.246 &  13.263 \ppm 0.10 &  12.054 \ppm 0.11 &  13.522 \ppm 0.08 &  12.147 \ppm 0.06 & $ 47^{+  36}_{-  21}$ \\ 
27 & 23 & 281219  & J175740.8-311443         & 17574089-3114437  & 1.55 &   0.614 &  14.736 &  12.939 \ppm 0.08 &  11.469 \ppm 0.08 &  12.759 \ppm 0.03 &  11.445 \ppm 0.02 & $ 49^{+  37}_{-  21}$ \\ 
28 & 25 & 327263  & J175441.4-331626         & 17544144-3316268  & 3.72 &   0.377 &  13.145 &   9.841 \ppm 0.05 &   8.183 \ppm 0.04 &   9.860 \ppm 0.02 &   8.221 \ppm 0.02 & $  5^{+   4}_{-   2}$ \\ 
29$^{\star,\dagger}$ & 25 & 592433  & J175443.1-323213         & 17544314-3232134  & 0.61 &    0.044 &  12.052 &  10.580 \ppm 0.06 &   9.712 \ppm 0.06 &  10.626 \ppm 0.03 &   9.768 \ppm 0.02 & $ 29^{+  22}_{-  13}$ \\ 
30 & 30 & 152246  & J180107.1-283547         & 18010706-2835458  & 1.86 &  1.947 &  11.998 &  10.486 \ppm 0.06 &   9.571 \ppm 0.07 &  10.410 \ppm 0.02 &   9.546 \ppm 0.03 & $ 25^{+  19}_{-  11}$ \\ 
31 & 34 & 159315  & J175752.2-285845         & 17575241-2858464  & 1.90 &  2.373 &  13.960 &  11.990 \ppm 0.08 &     &  12.170 \ppm 0.04 &  11.068 \ppm 0.04 & $ 27^{+  20}_{-  11}$ \\ 
32 & 34 & 465798  & J175803.7-284425         & 17580385-2844256  & 1.72 &  1.397 &  12.405 &  10.948 \ppm 0.06 &  10.183 \ppm 0.06 &  11.009 \ppm 0.03 &  10.218 \ppm 0.03 & $ 36^{+  27}_{-  16}$ \\ 
33 & 35 & 346292  & J180417.9-274610         & 18041796-2746096  & 2.94 &   0.888 &  13.873 &  11.281 \ppm 0.07 &     &  11.364 \ppm 0.02 &   9.880 \ppm 0.02 & $ 13^{+  10}_{-   5}$ \\ 
34 & 36 & 752090  & J180547.5-275834         & 18054752-2758351  & 1.14 &   0.735 &  14.186 &  12.505 \ppm 0.08 &  11.279 \ppm 0.08 &  12.545 \ppm 0.06 &  11.356 \ppm 0.04 & $ 46^{+  35}_{-  20}$ \\ 
35 & 37 & 341121  & J175233.7-302209         & 17523369-3022098  & 1.14 &   0.754 &  14.061 &  12.163 \ppm 0.11 &  10.943 \ppm 0.10 &  12.117 \ppm 0.03 &  10.946 \ppm 0.03 & $ 31^{+  24}_{-  13}$ \\ 
36 & 37 & 441811  & J175239.7-294552         & 17523977-2945521  & 3.19 &   0.499 &  13.655 &  11.944 \ppm 0.07 &  10.701 \ppm 0.07 &  12.061 \ppm 0.06 &  10.797 \ppm 0.05 & $ 34^{+  26}_{-  15}$ \\ 
37 & 37 & 514546  & J175249.8-301735         & 17524989-3017358  & 2.97 &   0.655 &  14.247 &  12.273 \ppm 0.07 &  10.958 \ppm 0.07 &  12.247 \ppm 0.05 &  10.918 \ppm 0.04 & $ 30^{+  23}_{-  13}$ \\ 
38 & 38 & 190202  & J180120.0-302352         & 18012009-3023526  & 2.74 &   0.643 &  13.770 &  11.906 \ppm 0.06 &  10.488 \ppm 0.07 &  11.894 \ppm 0.04 &  10.570 \ppm 0.03 & $ 28^{+  22}_{-  12}$ \\ 
39 & 39 & 309705  & J175531.5-294638         & 17553162-2946396  & 0.40 &  1.548 &  14.874 &  12.901 \ppm 0.08 &  11.548 \ppm 0.10 &  12.150 \ppm   &  10.809 \ppm   & $ 40^{+  31}_{-  18}$ \\ 
40$^{\star,\dagger}$ & 41 & 47647  & J175149.2-331543         & 17514924-3315428  & 0.47 &   0.616 &  13.830 &  12.018 \ppm 0.06 &  10.650 \ppm 0.07 &  12.027 \ppm 0.03 &  10.729 \ppm 0.02 & $ 31^{+  24}_{-  14}$ \\ 
41 & 41 & 353119  & J175221.2-331444         & 17522121-3314439  & 3.58 &   0.727 &  14.063 &  11.972 \ppm 0.11 &  10.714 \ppm 0.10 &  11.990 \ppm 0.05 &  10.554 \ppm   & $ 24^{+  18}_{-  10}$ \\ 
42 & 44 & 270634  & J174950.1-300004         & 17495008-3000044  & 3.01 &   0.887 &  14.002 &  10.915 \ppm 0.07 &   8.811 \ppm 0.06 &  10.792 \ppm 0.03 &   8.693 \ppm 0.03 & $  9^{+   7}_{-   4}$ \\ 
43$^{\star,\dagger}$ & 46 & 44548  & J180408.7-301305         & 18040875-3013055  & 0.62 &   0.101 &  14.128 &  12.485 \ppm 0.08 &  11.577 \ppm 0.13 &  12.480 \ppm 0.02 &  11.596 \ppm 0.02 & $ 48^{+  36}_{-  21}$ \\ 
44$^{\star,\dagger}$ & 46 & 335077  & J180454.0-301436         & 18045398-3014368  & 0.69 &   0.897 &  13.634 &  11.997 \ppm 0.09 &  10.983 \ppm 0.11 &  12.090 \ppm 0.04 &  11.023 \ppm 0.03 & $ 38^{+  29}_{-  17}$ \\ 

\end{tabular}

\label{tab:Mstars2MASS}
\end{sidewaystable}


\normalsize
\twocolumn



\begin{thebibliography}{38}
\providecommand{\natexlab}[1]{#1}
\providecommand{\url}[1]{\texttt{#1}}
\providecommand{\urlprefix}{URL }

\bibitem[{{Alcock} et~al.(2000){Alcock}, {Allsman}, {Alves}, {Axelrod},
  {Becker}, {Bennett}, {Cook}, {Drake} et~al.}]{2000ApJ...541..734A}
{Alcock}, C., {Allsman}, R.A., {Alves}, D.R., {Axelrod}, T.S., {Becker}, A.C.,
  {Bennett}, D.P., {Cook}, K.H., {Drake}, A.J., et~al., 2000, \apj, 541, 734

\bibitem[{{Alcock} et~al.(2001){Alcock}, {Allsman}, {Alves}, {Axelrod},
  {Becker}, {Bennett}, {Cook}, {Drake} et~al.}]{2001ApJ...562..337A}
{Alcock}, C., {Allsman}, R.A., {Alves}, D.R., {Axelrod}, T.S., {Becker}, A.C.,
  {Bennett}, D.P., {Cook}, K.H., {Drake}, A.J., et~al., 2001, \apj, 562, 337

\bibitem[{{Aubourg} et~al.(1993){Aubourg}, {Bareyre}, {Brehin}, {Gros},
  {Lachieze-Rey}, {Laurent}, {Lesquoy}, {Magneville}
  et~al.}]{1993Natur.365..623A}
{Aubourg}, E., {Bareyre}, P., {Brehin}, S., {Gros}, M., {Lachieze-Rey}, M.,
  {Laurent}, B., {Lesquoy}, E., {Magneville}, C., et~al., 1993, \nat, 365, 623

\bibitem[{{Baraffe} et~al.(1998){Baraffe}, {Chabrier}, {Allard}, and
  {Hauschildt}}]{1998A&A...337..403B}
{Baraffe}, I., {Chabrier}, G., {Allard}, F., {Hauschildt}, P.H., 1998, \aap,
  337, 403

\bibitem[{{Bergeron} et~al.(1995){Bergeron}, {Wesemael}, and
  {Beauchamp}}]{1995PASP..107.1047B}
{Bergeron}, P., {Wesemael}, F., {Beauchamp}, A., 1995, \pasp, 107, 1047

\bibitem[{{Bond} et~al.(2001){Bond}, {Abe}, {Dodd}, {Hearnshaw}, {Honda},
  {Jugaku}, {Kilmartin}, {Marles} et~al.}]{2001MNRAS.327..868B}
{Bond}, I.A., {Abe}, F., {Dodd}, R.J., {Hearnshaw}, J.B., {Honda}, M.,
  {Jugaku}, J., {Kilmartin}, P.M., {Marles}, A., et~al., 2001, \mnras, 327, 868

\bibitem[{{Carpenter}(2001)}]{2001AJ....121.2851C}
{Carpenter}, J.M., 2001, \aj, 121, 2851

\bibitem[{{Cutri} et~al.(2003){Cutri}, {Skrutskie}, {van Dyk}, {Beichman},
  {Carpenter}, {Chester}, {Cambresy}, {Evans} et~al.}]{2003tmc..book.....C}
{Cutri}, R.M., {Skrutskie}, M.F., {van Dyk}, S., {Beichman}, C.A., {Carpenter},
  J.M., {Chester}, T., {Cambresy}, L., {Evans}, T., et~al., 2003, {2MASS All
  Sky Catalog of point sources.}, The IRSA 2MASS All-Sky Point Source Catalog,
  NASA/IPAC Infrared Science
  Archive.~http://irsa.ipac.caltech.edu/applications/Gator/

\bibitem[{{Cutri} et~al.(2005){Cutri}, {Skrutskie}, {van Dyk}, {Beichman},
  {Carpenter}, {Chester}, {Cambresy}, {Evans} et~al.}]{2MASSExplSupp05}
{Cutri}, R.M., {Skrutskie}, M.F., {van Dyk}, S., {Beichman}, C.A., {Carpenter},
  J.M., {Chester}, T., {Cambresy}, L., {Evans}, T., et~al., 2005, {Explanatory
  Supplement to the 2MASS All Sky Data Release and Extended Mission Products},
  Technical report, Caltech,
  http://www.ipac.caltech.edu/2mass/releases/allsky/doc/\-explsup.html

\bibitem[{{Dahn} et~al.(2002){Dahn}, {Harris}, {Vrba}, {Guetter}, {Canzian},
  {Henden}, {Levine}, {Luginbuhl} et~al.}]{2002AJ....124.1170D}
{Dahn}, C.C., {Harris}, H.C., {Vrba}, F.J., {Guetter}, H.H., {Canzian}, B.,
  {Henden}, A.A., {Levine}, S.E., {Luginbuhl}, C.B., et~al., 2002, \aj, 124,
  1170

\bibitem[{{Edvardsson} et~al.(1993){Edvardsson}, {Andersen}, {Gustafsson},
  {Lambert}, {Nissen}, and {Tomkin}}]{1993A&A...275..101E}
{Edvardsson}, B., {Andersen}, J., {Gustafsson}, B., {Lambert}, D.L., {Nissen},
  P.E., {Tomkin}, J., 1993, \aap, 275, 101

\bibitem[{{Epchtein} et~al.(1997){Epchtein}, {de Batz}, {Capoani},
  {Chevallier}, {Copet}, {Fouque}, {Lacombe}, {Le Bertre}
  et~al.}]{1997Msngr..87...27E}
{Epchtein}, N., {de Batz}, B., {Capoani}, L., {Chevallier}, L., {Copet}, E.,
  {Fouque}, P., {Lacombe}, F., {Le Bertre}, T., et~al., 1997, The Messenger,
  87, 27

\bibitem[{{Eyer} and {Wo{\'z}niak}(2001)}]{2001MNRAS.327..601E}
{Eyer}, L., {Wo{\'z}niak}, P.R., 2001, \mnras, 327, 601

\bibitem[{{Gliese} and {Jahrei{\ss}}(1991)}]{1991adc..rept.....G}
{Gliese}, W., {Jahrei{\ss}}, H., 1991, {Preliminary Version of the Third
  Catalogue of Nearby Stars}, Technical report

\bibitem[{{Henry} et~al.(1997){Henry}, {Ianna}, {Kirkpatrick}, and
  {Jahreiss}}]{1997AJ....114..388H}
{Henry}, T.J., {Ianna}, P.A., {Kirkpatrick}, J.D., {Jahreiss}, H., 1997, \aj,
  114, 388

\bibitem[{{Hoeg} et~al.(1997){Hoeg}, {B{\"a}ssgen}, {Bastian}, {Egret},
  {Fabricius}, {Gro{\ss}mann}, {Halbwachs}, {Makarov}
  et~al.}]{1997A&A...323L..57H}
{Hoeg}, E., {B{\"a}ssgen}, G., {Bastian}, U., {Egret}, D., {Fabricius}, C.,
  {Gro{\ss}mann}, V., {Halbwachs}, J.L., {Makarov}, V.V., et~al., 1997, \aap,
  323, L57

\bibitem[{{Ibukiyama} and {Arimoto}(2002)}]{2002A&A...394..927I}
{Ibukiyama}, A., {Arimoto}, N., 2002, \aap, 394, 927

\bibitem[{{Landolt}(1992)}]{1992AJ....104..372L}
{Landolt}, A.U., 1992, \aj, 104, 372

\bibitem[{{Lejeune} et~al.(1997){Lejeune}, {Cuisinier}, and
  {Buser}}]{1997A&AS..125..229L}
{Lejeune}, T., {Cuisinier}, F., {Buser}, R., 1997, \aaps, 125, 229

\bibitem[{{Mao} and {Paczy{\'n}ski}(2002)}]{2002MNRAS.337..895M}
{Mao}, S., {Paczy{\'n}ski}, B., 2002, \mnras, 337, 895

\bibitem[{{Perryman} and {ESA}(1997)}]{1997ESASP1200.....P}
{Perryman}, M.A.C., {ESA}, editors, 1997, {The HIPPARCOS and TYCHO catalogues.
  Astrometric and photometric star catalogues derived from the ESA HIPPARCOS
  Space Astrometry Mission}, volume 1200 of ESA Special Publication

\bibitem[{{Perryman} et~al.(1997){Perryman}, {Lindegren}, {Kovalevsky}, {Hoeg},
  {Bastian}, {Bernacca}, {Cr{\'e}z{\'e}}, {Donati}
  et~al.}]{1997A&A...323L..49P}
{Perryman}, M.A.C., {Lindegren}, L., {Kovalevsky}, J., {Hoeg}, E., {Bastian},
  U., {Bernacca}, P.L., {Cr{\'e}z{\'e}}, M., {Donati}, F., et~al., 1997, \aap,
  323, L49

\bibitem[{{Phan-Bao} et~al.(2001){Phan-Bao}, {Guibert}, {Crifo}, {Delfosse},
  {Forveille}, {Borsenberger}, {Epchtein}, {Fouqu{\'e}}
  et~al.}]{2001A&A...380..590P}
{Phan-Bao}, N., {Guibert}, J., {Crifo}, F., {Delfosse}, X., {Forveille}, T.,
  {Borsenberger}, J., {Epchtein}, N., {Fouqu{\'e}}, P., et~al., 2001, \aap,
  380, 590

\bibitem[{{Rattenbury} et~al.(2007{\natexlab{a}}){Rattenbury}, {Mao},
  {Debattista}, {Sumi}, {Gerhard}, and {de Lorenzi}}]{2007MNRAS.378.1165R}
{Rattenbury}, N.J., {Mao}, S., {Debattista}, V.P., {Sumi}, T., {Gerhard}, O.,
  {de Lorenzi}, F., 2007{\natexlab{a}}, \mnras, 378, 1165

\bibitem[{{Rattenbury} et~al.(2007{\natexlab{b}}){Rattenbury}, {Mao}, {Sumi},
  and {Smith}}]{2007MNRAS.378.1064R}
{Rattenbury}, N.J., {Mao}, S., {Sumi}, T., {Smith}, M.C., 2007{\natexlab{b}},
  \mnras, 378, 1064

\bibitem[{{Reid} et~al.(2003){Reid}, {Cruz}, {Laurie}, {Liebert}, {Dahn},
  {Harris}, {Guetter}, {Stone} et~al.}]{2003AJ....125..354R}
{Reid}, I.N., {Cruz}, K.L., {Laurie}, S.P., {Liebert}, J., {Dahn}, C.C.,
  {Harris}, H.C., {Guetter}, H.H., {Stone}, R.C., et~al., 2003, \aj, 125, 354

\bibitem[{{Reyl{\'e}} and {Robin}(2004)}]{2004A&A...421..643R}
{Reyl{\'e}}, C., {Robin}, A.C., 2004, \aap, 421, 643

\bibitem[{{Reyl{\'e}} et~al.(2002){Reyl{\'e}}, {Robin}, {Scholz}, and
  {Irwin}}]{2002A&A...390..491R}
{Reyl{\'e}}, C., {Robin}, A.C., {Scholz}, R.D., {Irwin}, M., 2002, \aap, 390,
  491

\bibitem[{{Robin} et~al.(2003){Robin}, {Reyl{\'e}}, {Derri{\`e}re}, and
  {Picaud}}]{2003A&A...409..523R}
{Robin}, A.C., {Reyl{\'e}}, C., {Derri{\`e}re}, S., {Picaud}, S., 2003, \aap,
  409, 523

\bibitem[{{Soszynski} et~al.(2002){Soszynski}, {Zebrun}, {Udalski}, {Wozniak},
  {Szymanski}, {Kubiak}, {Pietrzynski}, {Szewczyk}
  et~al.}]{2002AcA....52..143S}
{Soszynski}, I., {Zebrun}, K., {Udalski}, A., {Wozniak}, P.R., {Szymanski}, M.,
  {Kubiak}, M., {Pietrzynski}, G., {Szewczyk}, O., et~al., 2002, Acta
  Astronomica, 52, 143

\bibitem[{{Sumi} et~al.(2003{\natexlab{a}}){Sumi}, {Abe}, {Bond}, {Dodd},
  {Hearnshaw}, {Honda}, {Honma}, {Kan-ya} et~al.}]{2003ApJ...591..204S}
{Sumi}, T., {Abe}, F., {Bond}, I.A., {Dodd}, R.J., {Hearnshaw}, J.B., {Honda},
  M., {Honma}, M., {Kan-ya}, Y., et~al., 2003{\natexlab{a}}, \apj, 591, 204

\bibitem[{{Sumi} et~al.(2003{\natexlab{b}}){Sumi}, {Eyer}, and
  {Wo{\'z}niak}}]{2003MNRAS.340.1346S}
{Sumi}, T., {Eyer}, L., {Wo{\'z}niak}, P.R., 2003{\natexlab{b}}, \mnras, 340,
  1346

\bibitem[{{Sumi} et~al.(2004){Sumi}, {Wu}, {Udalski}, {Szyma{\'n}ski},
  {Kubiak}, {Pietrzy{\'n}ski}, {Soszy{\'n}ski}, {Wo{\'z}niak}
  et~al.}]{2004MNRAS.348.1439S}
{Sumi}, T., {Wu}, X., {Udalski}, A., {Szyma{\'n}ski}, M., {Kubiak}, M.,
  {Pietrzy{\'n}ski}, G., {Soszy{\'n}ski}, I., {Wo{\'z}niak}, P., et~al., 2004,
  \mnras, 348, 1439

\bibitem[{{The Denis Consortium}(2005)}]{2005yCat....102002T}
{The Denis Consortium}, 2005, VizieR Online Data Catalog, 1, 2002

\bibitem[{{Udalski} et~al.(1997){Udalski}, {Kubiak}, and
  {Szymanski}}]{1997AcA....47..319U}
{Udalski}, A., {Kubiak}, M., {Szymanski}, M., 1997, Acta Astronomica, 47, 319

\bibitem[{{Udalski} et~al.(2002){Udalski}, {Szymanski}, {Kubiak},
  {Pietrzynski}, {Soszynski}, {Wozniak}, {Zebrun}, {Szewczyk}
  et~al.}]{2002AcA....52..217U}
{Udalski}, A., {Szymanski}, M., {Kubiak}, M., {Pietrzynski}, G., {Soszynski},
  I., {Wozniak}, P., {Zebrun}, K., {Szewczyk}, O., et~al., 2002, Acta
  Astronomica, 52, 217

\bibitem[{{Udalski} et~al.(2000){Udalski}, {Zebrun}, {Szymanski}, {Kubiak},
  {Pietrzynski}, {Soszynski}, and {Wozniak}}]{2000AcA....50....1U}
{Udalski}, A., {Zebrun}, K., {Szymanski}, M., {Kubiak}, M., {Pietrzynski}, G.,
  {Soszynski}, I., {Wozniak}, P., 2000, Acta Astronomica, 50, 1

\bibitem[{{Woolley}(1970)}]{1970ROAn....5.....W}
{Woolley}, R.v.d.R., 1970, Royal Observatory Annals, 5

\end{thebibliography}

\end{document}